\documentclass[12pt]{article}

\usepackage{array} 
\usepackage{amssymb}
\usepackage{graphics,graphpap}
\usepackage{graphicx}
\usepackage{color}
\usepackage{graphicx}
\usepackage{dcolumn}
\usepackage{epsfig}
\usepackage{epstopdf}
\usepackage{bm}
\usepackage{amsmath}
\usepackage{amsfonts}
\usepackage{textcomp}
\usepackage{setspace}

\newcommand{\beq}{\begin{eqnarray}}
\newcommand{\eeq}{\end{eqnarray}}

\newcommand{\bmp}{\noindent\begin{minipage}{16cm}}
\newcommand{\emp}{\end{minipage}\vskip 7mm} 

\def\drawbox#1#2{\hrule height#2pt
        \hbox{\vrule width#2pt height#1pt \kern#1pt
              \vrule width#2pt}
              \hrule height#2pt}

\def\Asym#1#2{\vcenter{\vbox{\drawbox{#1}{#2}
              \kern-#2pt 
              \drawbox{#1}{#2}}}}

\def\simge{\mathrel{%
   \rlap{\raise 0.511ex \hbox{$>$}}{\lower 0.511ex \hbox{$\sim$}}}}

\def\simle{\mathrel{
   \rlap{\raise 0.511ex \hbox{$<$}}{\lower 0.511ex \hbox{$\sim$}}}}

\def\s#1{\setbox0=\hbox{$#1$}%
\rlap{\ifdim\wd0>.7em\kern.22\wd0\else\kern.1\wd0\fi /}#1}

\begin{document}

\begin{titlepage}
\title{\vspace*{-2.0cm}
\bf\Large
Warm Dark Matter from keVins \\[5mm]\ }

\author{
Stephen F.\ King$^a$\thanks{email: \tt S.F.King@soton.ac.uk}~~and
Alexander Merle$^b$\thanks{email: \tt amerle@kth.se}
\\ \\
{\normalsize $^a$ \it School of Physics and Astronomy, University of Southampton,}\\
{\normalsize \it Southampton, SO17 1BJ, United Kingdom}\\
\\
{\normalsize $^b$ \it Department of Theoretical Physics, School of Engineering Sciences,}\\
{\normalsize \it KTH Royal Institute of Technology -- AlbaNova University Center,}\\
{\normalsize \it Roslagstullsbacken 21, 106 91 Stockholm, Sweden}
}
\date{\today}
\maketitle
\thispagestyle{empty}

\begin{abstract}
\noindent
We propose a simple model for Warm Dark Matter (WDM) in which two fermions are added to the Standard Model: (quasi-) stable ``keVins'' (keV inert fermions) which account for WDM and their unstable brothers,  the ``GeVins'' (GeV inert fermions), both of which carry zero electric charge and lepton number, and are (approximately) ``inert'', in the sense that their only interactions are via suppressed couplings to the $Z$. We consider scenarios in which stable keVins are thermally produced and their abundance is subsequently diluted by entropy production from the decays of the heavier unstable GeVins. This mechanism could be implemented in a wide variety of models, including $E_6$ inspired supersymmetric models or models involving sterile neutrinos.
\end{abstract}

\end{titlepage}

\section{\label{sec:intro}Introduction}

One of the biggest questions in contemporary astroparticle physics is about the nature of Dark Matter (DM), which makes up more than 20\% of the energy content of the current Universe~\cite{Komatsu:2010fb}. When considering particle physics motivated candidates for a DM particle, one distinguishes between \emph{cold} (CDM) and \emph{hot} Dark Matter (HDM), depending on if the freeze-out of the corresponding particles happened when they were already non-relativistic or when they still had relativistic velocities. Since a possible DM candidate particle should be electrically neutral, not strongly interacting, and (quasi-) stable, the only possibility within the Standard Model (SM) of particle physics would be a light neutrino. Due to strong experimental constraints (from single-$\beta$ decay experiments~\cite{Lobashev:2000vb,Kraus:2004zw}, double-$\beta$ decay experiments~\cite{KlapdorKleingrothaus:2000sn,Andreotti:2010vj}, and cosmological bounds~\cite{Komatsu:2010fb}), the neutrino mass should be less than about $1$~eV, making it a perfect HDM candidate. Unfortunately, HDM is excluded as the dominant DM component due to problems with structure fomation~\cite{Abazajian:2004zh,dePutter:2012sh}. On the other hand, many theories beyond the SM predict so-called \emph{WIMPs} (weakly interacting massive particles), which would be very suitable CDM candidates.

However, recently the intermediate case of \emph{warm} Dark Matter (WDM) has attracted considerable attention, in particular from the structure formation side. After pioneering works~\cite{Bode:2000gq,Hansen:2001zv} more than ten years ago, several simulations have been performed (see, e.g., Refs.~\cite{Boyarsky:2008xj,Lovell:2011rd}), in particular to study the small scale structure~\cite{Boyanovsky:2010pw,Boyanovsky:2010sv,VillaescusaNavarro:2010qy} induced by WDM, in order to maybe be able to solve the typical CDM problem of predicting many more dwarf satellite galaxies than observed. In addition, model-independent analyses~\cite{deVega:2009ku,deVega:2010yk} and surveys such as ALFALFA~\cite{Papastergis:2011xe} seem to point into the direction of WDM. In general we shall refer to the candidates responsible for WDM as  ``\emph{keVins}'' (\underline{keV}-mass \underline{in}ert particle\underline{s}). Just like WIMPs, keVins also share certain common properties, which is why it makes sense to identify them as a general class of WDM candidates that can be studied both in a general context and in a specific model.

Probably the leading particle physics candidate for a keVin that has been studied most frequently is a sterile neutrino with a mass of a few keV. Frameworks in which such a particle can appear are, e.g., the so-called $\nu$MSM~\cite{Asaka:2005an}, the left-right ($LR$)-symmetric framework~\cite{Bezrukov:2009th}, the scotogenic framework~\cite{Sierra:2008wj,Gelmini:2009xd}, or they could appear as composite neutrinos~\cite{Grossman:2010iq}. When studying keV sterile neutrinos as DM candidates, there are two main tasks, namely finding a suitable production mechanism to generate the correct DM abundance and explaining the very existence of the keV mass scale. Possible production mechanisms are, for example, non-resonant~\cite{Dodelson:1993je} and resonant~\cite{Shi:1998km} non-thermal production or thermal production with subsequent entropy dilution~\cite{Scherrer:1984fd}, which all have been applied to the keV sterile neutrino case~\cite{Bezrukov:2009th,Asaka:2006ek,Asaka:2006rw,Asaka:2006nq,Laine:2008pg,Wu:2009yr,Shaposhnikov:2006xi,Bezrukov:2012as}. To motivate the existence of a keV scale in the first place, the possibilities include the \emph{split seesaw} mechanism~\cite{Kusenko:2010ik,Adulpravitchai:2011rq}, flavour symmetries~\cite{Shaposhnikov:2006nn,Lindner:2010wr,Araki:2011zg}, the \emph{Froggatt-Nielsen} mechanism~\cite{Barry:2011wb,Merle:2011yv,Barry:2011fp}, or extensions of the seesaw mechanism~\cite{Zhang:2011vh}. See, e.g., Refs.~\cite{Chen:2011ai,Merle:2012ya,Abazajian:2012ys} for more general discussions. However, apart from keV neutrinos, there could also be other particles playing the role of keVins. The cases studied in the literature include the gravitino~\cite{Gorbunov:2008ui,Jedamzik:2005sx,Baltz:2001rq}, the KK-graviton~\cite{Jedamzik:2005sx}, singlinos~\cite{McDonald:2008ua}, axinos~\cite{Jedamzik:2005sx}, axion-axino mixed DM~\cite{Baer:2008yd}, or majorons~\cite{Lattanzi:2007ux,Frigerio:2011in}.

In this paper we propose a new and very simple model of WDM based on adding a pair of electrically neutral fermions to the Standard Model: a stable keVin, $\chi_1$, accompanied by an unstable partner, the ``GeVin'' $\chi_2$, with a mass in the GeV range. Both particles interact only by suppressed couplings to the SM gauge boson $Z$. These suppressed couplings may arise, e.g., from a small mixing effect in more sophisticated models. We show that the correct Dark Matter relic abundance can be obtained by thermal overproduction of $\chi_1$, which is subsequently diluted by the production of additional entropy from the decay of $\chi_2$, while satisfying constraints from structure formation and big bang nucleosynthesis. Such a mechanism could be applied to a wide variety of models, for example $E_6$ inspired supersymmetric models~\cite{King:2005jy} or models involving sterile neutrinos~\cite{Bezrukov:2009th}. However, we stress that it is also possible to have keVins without GeVins, with the keVin being produced non-thermally due to its feeble interactions, which is the traditional mechanism for warm Dark Matter from keV sterile neutrinos. 

This paper is organized as follows. In Sec.~\ref{sec:keVins}, we first introduce the general framework of keVins and perform a detailed semi-numerical analysis of how to generate the correct DM abundance using thermal (over-) production of the keVins in the early Universe. This is achieved by suppressed couplings of the keVins to the $Z$--boson. The abundance is later on diluted by additional entropy produced in decays of a heavier electrically neutral particle, the GeVin $\chi_2$. We also identify several benchmark points for the five independent parameters, which can then be used to check specific models. Exactly that will be done in Sec.~\ref{sec:Concrete-keVins}, where we first present a situation in the framework of the $E_6$SSM in which keVins naturally arise. After that, we also give a short discussion on how to extend our consideration to the case of keV sterile neutrinos in a $LR$-symmetric framework. These examples serve as guidelines on how to apply our considerations, but they are certainly not the only examples one could find. Finally, in Sec.~\ref{sec:conclusions}, we will summarize our findings and conclude.

\section{\label{sec:keVins} keVins and GeVins}

In this section, we introduce a simple model of WDM in which two fermions are added to the Standard Model: keVins and GeVins, both of which carry zero electric charge and lepton number, and only interact via suppressed couplings to the $Z$. Later on, in Sec.~\ref{sec:Concrete-keVins}, we will apply this mechanism to other examples, e.g.\ within the $E_6$SSM or to models with sterile neutrinos. We want to stress, however, that such examples do not play any decisive role, and that other examples could easily be found which also involve the simple mechanism discussed in this section.

\subsection{\label{sec:keVins_basic}The $Z$ couplings}

The starting point of the model is to add a pair of electrically neutral fermions to the Standard Model: a stable keVin, $\chi_1$, accompanied by an unstable GeVin, $\chi_2$. Both these fields are assumed to have suppressed couplings to the SM gauge boson $Z$. This may be achieved by having a singlet fermion with small mixing to other fermions,  
as in the $E_6$ models discussed later in the paper. However, here we simply assume suppressed couplings of keVins and GeVins to the $Z$. Note that the electric charges of $\chi_i$ are strictly zero. In addition, we consider the possibility for the $Z$ to also couple in a flavour-changing way to $\chi_1$ and $\chi_2$, allowing decays such as $\chi_2 \rightarrow \chi_1 + Z^*$,  $Z^* \rightarrow f\overline{f} $, where $f$ is any kinematically accessible SM fermion.

The masses of $\chi_{1,2}$ are denoted as $M_{1,2}$. What we have in mind is to have masses $M_1 \sim \mathcal{O}(1-100~{\rm keV})$, while $M_2 \gtrsim \mathcal{O}(1-100~{\rm GeV})$ is considerably heavier. Hence, we have arrived at a framework where we have one particle with a keV mass that is stabilized by some additional quantum number, for example $R$-parity in supersymmetric models. This particle is mainly a SM singlet fermion, which is why the name \emph{keVin} (\underline{keV}-mass \underline{in}ert particle) is appropriate. Furthermore, the keVin is the WDM candidate. 

Before we give an estimate of the DM-abundance, which turns out to be easily and naturally reproducible in this simple framework, we want to comment on the bounds arising from the $Z$--boson decay width: both our singlets, the keVin $\chi_1$ as well as its heavier brother, the GeVin $\chi_2$, will have suppressed couplings to the SM $Z$--bosons, given by the small parameters $\epsilon_i$ and $\delta$. We assume the following interaction Lagrangians:\footnote{Note that we assume for the $\chi_i$ a pure axial vector coupling to the SM $Z$--boson, for simplicity. Our considerations would, however, not be considerably altered if any type of mixed vector/axial vector coupling was assumed, and the reader is invited to repeat our analysis for such a case.}
\begin{itemize}

\item $Z$--$\chi_i$--$\chi_i$ ($i=1,2$):
\begin{equation}
 \mathcal{L}_{ii} = g Z_{\mu}\epsilon_i^2 \overline{\chi_i} \gamma^\mu \gamma_5 \chi_i,
 \label{eq:Lii}
\end{equation}

\item $Z$--$\chi_1$--$\chi_2$:
\begin{equation}
 \mathcal{L}_{12} = g  \epsilon_1 \epsilon_2 \delta Z_{\mu} \overline{\chi_1} \gamma^\mu \gamma_5 \chi_2 + h.c.,
 \label{eq:L12}
\end{equation}

\end{itemize}
where $g=0.653$ is the usual $SU(2)$ gauge coupling, and $\epsilon_i$ are the suppressed couplings $\chi_i$ to the $Z$. Note that, if the masses $M_{1,2}$ are smaller than $M_Z /2$, they would contribute to the invisible $Z$--boson decay width, whose value is given by $\Gamma_{Z,\rm invisible}=(499.0\pm 1.5)$~MeV~\cite{Nakamura:2010zzi}. Since this value is more or less exactly what is obtained by considering the decay of $Z$ into pairs of the three known active neutrino flavours, the corresponding decay width into inert fermions,
\begin{equation}
 \Gamma_\chi \equiv \Gamma (Z \to \chi_1 \chi_1) + \Gamma (Z \to \chi_1 \chi_2) + \Gamma (Z \to \chi_2 \chi_2) = \frac{g^2 M_Z}{24 \pi} \left( \epsilon_1^4 + \epsilon_2^4 + 2  \epsilon_{12}^2 \right),
 \label{eq:Z-decay}
\end{equation}
where $ \epsilon_{12} \equiv \epsilon_1 \epsilon_2 \delta$, must be smaller than roughly the uncertainty of the invisible decay width, $\Gamma_\chi \lesssim 1.5$~MeV. Assuming that a certain $\epsilon_i$ dominates, this will turn into a bound of $\epsilon_i \lesssim 0.23$.

This leaves us with only five decisive parameters: $M_{1,2}$, $\epsilon_{1,2}$, and $\delta$.

\subsection{\label{sec:keVins_FO}Freeze-out regions}

\begin{figure}[t]
\centering
\includegraphics[width=7.6cm]{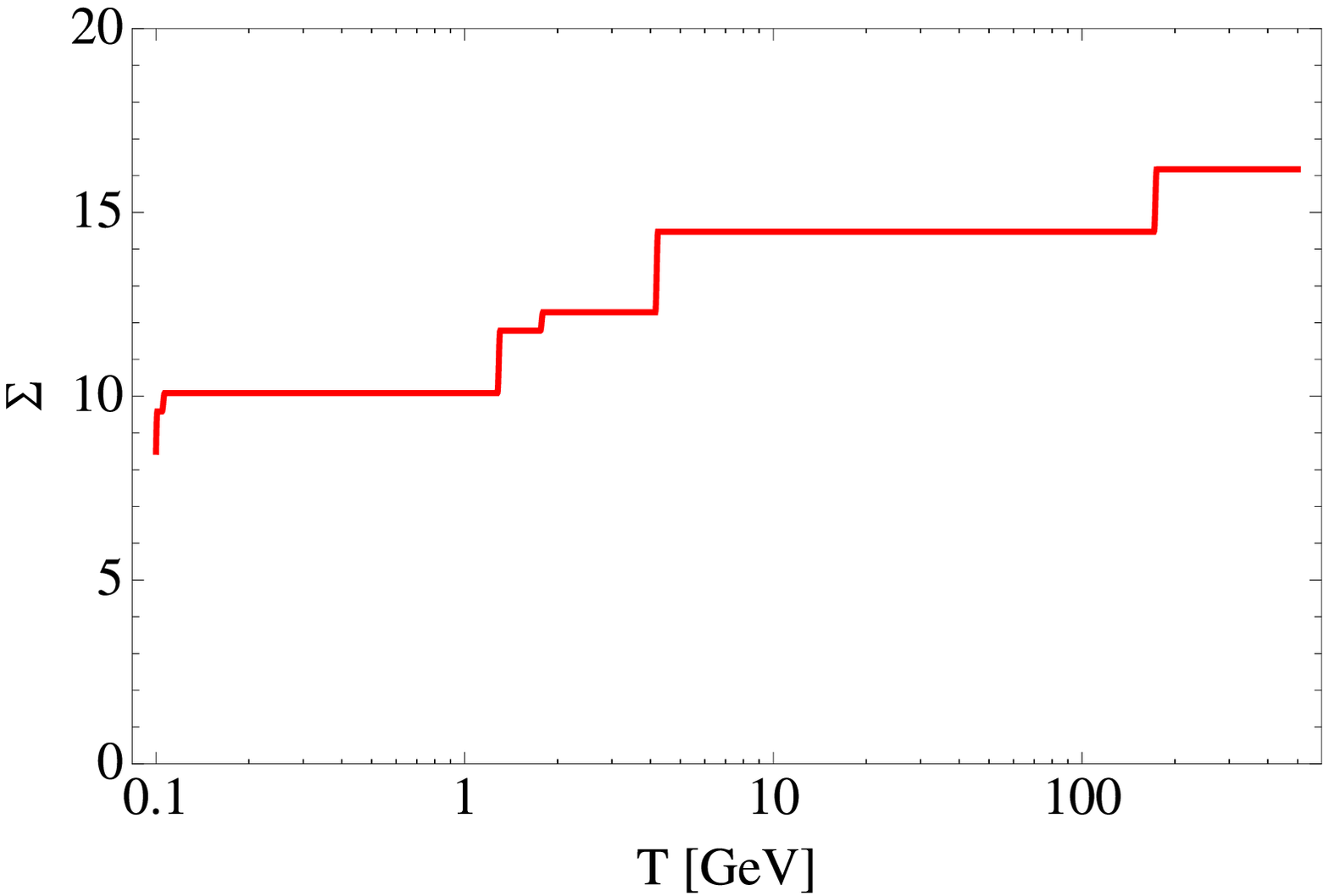}
\includegraphics[width=8.0cm]{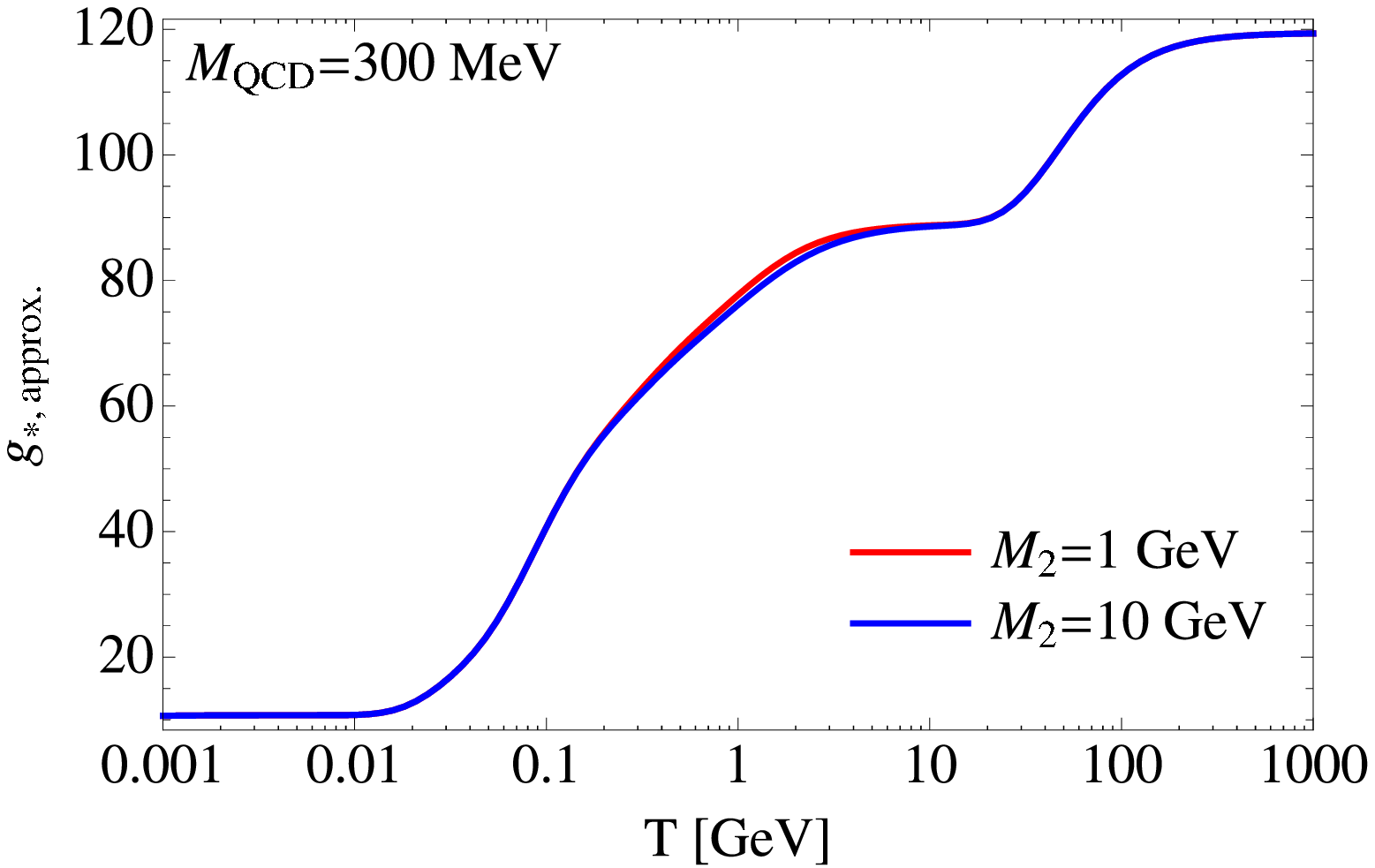}
\caption{\label{fig:Sig_g} The quantities $\Sigma$ and $g_*$ as functions of the temperature. Thereby, $g_*$ includes the contributions from the two light fermions that are present in our framework in addition to the SM particle content, i.e., the keVin $\chi_1$ and the GeVin $\chi_2$, making it different from its SM counterpart. Note that the biggest ``jump'' in $g_*$ is at the QCD scale, where quarks are bound into nucleons. We have taken the example value of $M_{\rm QCD} = 300$~MeV for definiteness (see Ref.~\cite{Nakamura:2010zzi} for more details on that point).}
\end{figure}

The crucial values to be determined are the freeze-out temperatures of the two particles under consideration, $\chi_1$ and $\chi_2$. Therefore, we consider the annihilations of a $\chi_i$--$\chi_i$ pair into either a SM fermion-antifermion pair, or into a pair of $W$--bosons, by a $Z$--boson mediator, depending on the temperature and initial state mass. Note that we neglect meson final states since they will, in the relevant parameter regions, either be negligible or kinematically not accessible. The corresponding spin-averaged matrix elements $\overline{|\mathcal{M}|^2_f}$ and $\overline{|\mathcal{M}|^2_W}$ for annihilation into fermion pairs or $W$--bosons, respectively, are given by
\begin{eqnarray}
 \overline{|\mathcal{M}|^2_f} &=& \frac{4 g^4 \epsilon_i^4}{c_W^2 \left[(s-M_Z^2)^2 + M_Z^2 \Gamma_Z^2 \right]} \Sigma \left[ (k_1 p_1) (k_2 p_2) + (k_1 p_2) (k_2 p_1) - M_i^2 (p_1 p_2) \right],\nonumber \\
 \overline{|\mathcal{M}|^2_W} &=& \frac{2 g^4 \epsilon_i^4 c_W^2}{(s-M_Z^2)^2 + M_Z^2 \Gamma_Z^2} \left[ k_1 (p_1 - p_2) \cdot k_2 (p_1 - p_2) + (p_1 p_2) (k_1 k_2 - M_i^2) \right].
 \label{eq:MEs}
\end{eqnarray}
Here, $k_i$ ($p_i$) are the initial (final) state momenta, $c_W\equiv \cos \theta_W$ is the cosine of the Weinberg angle, $\Gamma_Z = (2.4952 \pm 0.0023)$~GeV is the $Z$--boson total decay width, $\sqrt{s}$ is the center-of-mass energy, and $\Sigma = \sum_f \left( A_f^2 + B_f^2 \right)$ with $A_f$ and $B_f$ given in terms of SM fermion vector and axial vector couplings as $A_f = g_V^f + g_A^f$ and $B_f = g_V^f - g_A^f$, where the sum runs over all fermions that are kinematically accessible. We have plotted $\Sigma$ as function of the temperature in Fig.~\ref{fig:Sig_g}, left panel.

The two limiting cases for which analytical solutions to the Boltzmann equation exist are those of relativistic freeze-out (\emph{hot thermal relic}) and of non-relativistic freeze-out (\emph{cold thermal relic}). For both these regions, we have first computed the corresponding approximations of the expressions in Eq.~\eqref{eq:MEs}, and have then performed the thermal average. In case of relativistic freeze-out, this results in
\begin{equation}
\langle \sigma v \rangle_{\rm rel.}^f = \frac{g^4 \epsilon_i^4 \Sigma}{24 \pi c_W^2} F(T)\ \ {\rm and}\ \ \langle \sigma v \rangle_{\rm rel.}^W = \frac{g^4 \epsilon_i^4 c_W^2}{24 \pi} F(T),
 \label{eq:sv_rel}
\end{equation}
where $F(T)\simeq A\left( \frac{s}{(s-M_Z^2)^2} \right) / A(1)$, and the function $A(X)$ is given by
\begin{equation}
 A(X) = \int\limits_{s=0}^\infty X s^{3/2} K_1 \left( \frac{\sqrt{s}}{T} \right).
 \label{eq:A(X)}
\end{equation}
Note that $A(1) = 32\ T^5$ can be calculated analytically, while $A\left( \frac{s}{(s-M_Z^2)^2} \right)$ is evaluated numerically. In the non-relativistic limit, in turn, we obtain\footnote{Note that, due to having two Majorana fermions in the initial state, the leading term in the non-relativistic approximation of the annihilation cross section is the $p$--wave, which is suppressed by a factor $v^2$ of the velocity. This translates into a factor $6 T / M_2$ in the thermal average. For a detailed discussion of the cases where such suppressions are present or not, see Ref.~\cite{Lindner:2010rr}.}
\begin{equation}
\langle \sigma v \rangle_{\rm non\text{-}rel.}^f = \frac{g^4 \epsilon_i^4 \Sigma}{2 \pi c_W^2}\cdot \frac{M_i T}{(s-M_Z^2)^2 + M_Z^2 \Gamma_Z^2}\ \ {\rm and}\ \ \langle \sigma v \rangle_{\rm non\text{-}rel.}^W = \frac{g^4 \epsilon_i^4 c_W^2}{4 \pi} \cdot \frac{M_i T}{(s-M_Z^2)^2 + M_Z^2 \Gamma_Z^2}.
 \label{eq:sv_non}
\end{equation}
Furthermore, we need the number densities for the relativistic and non-relativistic cases, which are given by
\begin{equation}
n_{\rm rel.} = \frac{3}{4} g_\chi \frac{\zeta(3)}{\pi^2}\ T^3\ \ {\rm and}\ \ n_{\rm non\text{-}rel.} = g_\chi \left( \frac{M_i T}{2 \pi} \right)^{3/2}\ e^{-M_i/T},
 \label{eq:n_den}
\end{equation}
where $g_\chi = 2$ is the number of internal degrees of freedom of the fermion $\chi_i$, $\zeta(x)$ is the Riemann zeta function, and we have neglected the chemical potential in the non-relativistic case.

\begin{figure}[t]
\centering
\includegraphics[width=13cm]{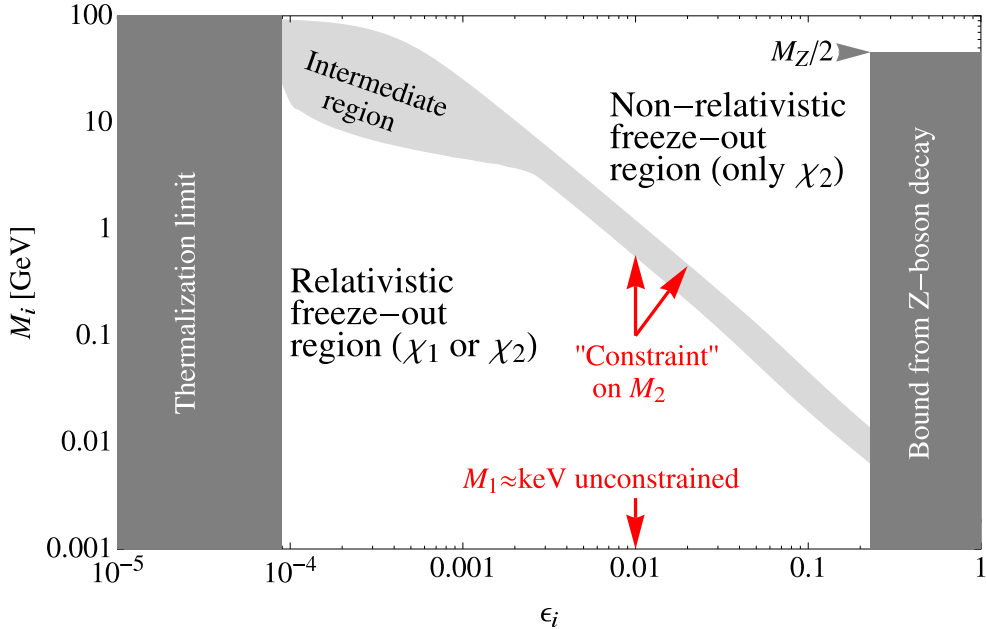}
\caption{\label{fig:FO-regions} Regions of relativistic and non-relativistic freeze-out. The intermediate region is marked by the light gray band. (See text for further explanations.)}
\end{figure}

The freeze-out temperature is then determined by setting the product $n \langle \sigma v \rangle$ equal to the Hubble rate $H = \frac{2 \pi^{3/2}}{3 \sqrt{5} M_P} \sqrt{g_*}\ T^2$,\footnote{There is no need to perform a more sophisticated procedure to determine the freeze-out temperature, since we will in any case use some other approximations, in particular when solving the Boltzmann equation.} where $M_P = 1.22\cdot 10^{19}$~GeV is the Planck mass and $g_*$ is the effective number of relativistic degrees of freedom (cf.\ Fig.~\ref{fig:Sig_g}, right panel). Finally, by setting the freeze-out temperature $T_{\rm FO}$ in the respective case equal to the mass $M_i$, we can obtain the (approximate) range of validity for both limiting cases. The result is displayed in Fig.~\ref{fig:FO-regions}. In the two limiting regions, it is possible to obtain an analytical estimate for the DM number density normalized by the entropy density, $Y_{\chi_i}=n_i/s$, while in the intermediate region one would need to numerically solve the Boltzmann equation to obtain $Y_{\chi_i}$. Such a numerical solution is, however, beyond the scope of this paper, and we will leave that more elaborate investigation for future work. Hence, in our case, we obtain an additional ``constraint'' on the masses. This does not play a role for $M_1\sim$~keV, which will always be in the relativistic freeze-out region. The heavier particle $\chi_2$, in turn, could freeze-out relativistically or non-relativistically, depending on the exact combination of mass $M_2$ and coupling suppression $\epsilon_2$.

Note that, for a too strong suppression $\epsilon_i$ in the annihilation cross section, the product $n \langle \sigma v \rangle$ will always be smaller than $H$, i.e., the particle never enters thermal equilibrium. In that case, thermal freeze-out is not possible, which is marked by the left gray rectangle in Fig.~\ref{fig:FO-regions}.\footnote{It may, however, be possible for very small couplings to produce the correct DM-abundance by the so-called \emph{freeze-in} instead. In that case, the corresponding particle is often called \emph{FIMP} (feebly interacting massive particle)~\cite{Hall:2009bx}.} For too large couplings, in turn, and a mass $M_i < M_Z /2$, there is the bound by the $Z$--boson decay width, marked by the right gray rectangle in the plot.

\subsection{\label{sec:keVins_analytical}Analytical estimates}

Before turning to a detailed analysis, it is useful to give some analytical estimates. The principle idea is the following: since the keVin $\chi_1$ freezes out while still being relativistic and since it has a mass $M_1$ of a few keV, its abundance will be much too large for it to play the role of the DM. However, if there is at some point after the freeze-out of $\chi_1$ a phase where the energy density of the Universe is dominated by the heavier fermion $\chi_2$, the out-of-equilibrium decays of the heavier particle (into relativistic final states, i.e., radiation) could produce a certain additional amount of entropy, parametrized by the factor $\mathcal{S}$, since the entropy of radiation is much higher than that of matter. This additional entropy delays the cooling of the Universe~\cite{Scherrer:1984fd} and it dilutes the natural abundance of $\chi_1$ by exactly that factor $\mathcal{S}$~\cite{Bezrukov:2009th,Asaka:2006ek}, which can correct its value in such a way to meet the observed region.\footnote{Note that this argumentation does not take into account the additional population of $\chi_1$ by the decays $\chi_2 \to \chi_1 f \overline{f}$ (and $\chi_2 \to \chi_1 W^- W^+$ for very large $M_2$). This should be a very good approximation: even though $\chi_2$ dominates the energy density of the Universe, $\rho_{\rm tot}\simeq \rho(\chi_2)$, at the time of its decay, its number density $n(\chi_2)=\rho(\chi_2)/M_2$ is much smaller than the number density of $\chi_1$, due to $M_2 \gg M_1$ and the late freeze-out. Thus, the main contribution of this decay to the DM abundance is indeed the dilution by entropy production.}

In that case, the final $\chi_1$ abundance will be given by~\cite{Bezrukov:2009th}
\begin{equation}
 \Omega_{\chi_1} h^2 = \frac{7.61\cdot 10^2}{\mathcal{S}}\ \frac{M_1}{10\ \rm keV}\ \frac{g_{\chi,\rm eff}}{g_{*s}(T_{\rm FO}^1)},
 \label{eq:1-abundance}
\end{equation}
where $g_{\chi,\rm eff} = \frac{3}{4} g_\chi$ and $g_{*s}(T_{\rm FO}^1) \simeq g_*(T_{\rm FO}^1)$. Here, $T_{\rm FO}^1$ denotes the freeze-out temperature of $\chi_1$. The factor $\mathcal{S}$ is given by the ratio between the entropy $S_f$ after most of the $\chi_2$ decays [i.e., for $t > \tau (\chi_2)$] and the entropy $S_i$ before the decays set in. This ratio is derived in detail in Ref.~\cite{Scherrer:1984fd}, and we will essentially make use of Eq.~(23c) therein:
\begin{equation}
 \mathcal{S} = \frac{S_f}{S_i} \simeq \left( 1 + 2.95 \left( \frac{2 \pi^2 g_*}{45} \right)^{1/3} \frac{(Y_{2,\infty} M_2)^{4/3}}{(M_P \Gamma_2)^{2/3}} \right)^{3/4},
 \label{eq:ent-prod}
\end{equation}
where we have approximated the weighted average of the effective number of relativistic degrees of freedom by the actual such number $g_*$ at the time under consideration. Furthermore, $Y_{2,\infty}$ is the final abundance of $\chi_2$ particles as obtained by the Boltzmann equation in the limit of small temperatures, and $\Gamma_2$ is the decay width of $\chi_2$. The decays under consideration involve a flavour changing coupling of the $Z$--boson, so that $\chi_2$ decays to $\chi_1$ and a fermion-antifermion pair. \footnote{Note that, in principle, there exists also a decay mode $\chi_2 \to 3 \chi_1$. However, this mode is suppressed by yet four more powers of $\epsilon_1$, so we neglect it here. In principle $\chi_2$ can also decay into $\chi_1$ plus a pair of $W$--bosons via the $Z$--$W$--$W$ coupling, 
however over most of the physical region of interest this decay is kinematically forbidden.} Taking the final states to be effectively massless, it is easy to derive the decay width:
\begin{equation}
 \Gamma (\chi_2 \to \chi_1 f \bar{f}) \simeq \frac{g^4 \xi^2 \Sigma}{1536 \pi^3 c_W^2} \cdot \frac{M_2^5}{(M_2^2 - M_Z^2)^2 + M_Z^2 \Gamma_Z^2}.
 \label{eq:f-decay}
\end{equation}

To obtain an easy estimate, we should first note that the natural value of the DM abundance, Eq.~\eqref{eq:1-abundance} with $\mathcal{S}\equiv 1$, is larger than the range allowed by WMAP-7~\cite{Komatsu:2010fb}, $\Omega_{\rm DM} h^2= 0.1120 \pm 0.0056$, by a factor much larger than 1. A large correction factor $\mathcal{S}$ is achieved most easily if $\chi_2$ freezes out while still being relativistic, where we have $Y_{2,\infty} = \frac{g_\chi}{2} \frac{135 \zeta(3)}{4 \pi^4 g_*}$, since in that case the number density of $\chi_2$ is not suppressed. Hence, in the region where we could potentially reproduce the correct abundance, one can neglect the first term in Eq.~\eqref{eq:ent-prod}~\cite{Bezrukov:2009th}, which allows us to write
\begin{equation}
 \mathcal{S} \approx 0.76\ \frac{g_\chi}{2} \frac{g_*^{1/4} M_2}{g_* \sqrt{\Gamma_2 M_P}},
 \label{eq:ent-prod_approx}
\end{equation}
where we have again approximated the averaged value of $g_*$ by an example value in the region of interest, since it is slowly varying anyway in most of the parameter space. In order to obtain the required decay width $\Gamma_{\rm req}$ needed to meet the correct relic abundance, we can combine Eqs.~\eqref{eq:1-abundance} and~\eqref{eq:ent-prod_approx}:
\begin{equation}
 \Gamma_{\rm req} \simeq 0.50\cdot 10^{-6}\ \left(\frac{g_*(T_{\rm FO}^1)}{g_*}\right)^2\ g^{1/2}_* \frac{M_2^2}{M_P} \left( \frac{\rm keV}{M_1} \right)^2.
 \label{eq:Gamma_req}
\end{equation}
In our case, we have $g_* = g_*(T_{\rm FO}^1)$ ($\nu_{e,\mu,\tau}$, $e^\pm$, and $\chi_1$ are relativistic)\footnote{Note that this leads to $g_* \approx 10$, which is the smallest possible value of $g_*$ before the inset of Big Bang Nucleosynthesis. However, if we choose a smaller coupling $\epsilon_1$ it may be that the keVin $\chi_1$ freezes out much earlier, at a higher temperature. If this freeze-out happens before the QCD phase transition, while the entropy producing decays of $\chi_2$ happen after the QCD transition, then the factor $[g_*(T_{\rm FO}^1)/g_*]^2$ leads to a significant boost of the RHS of Eq.~\eqref{eq:Gamma_req}, which could, for $\epsilon_1 = 0.001$, be as large as about 25. This would lead to an increase of the reheating temperature by a factor of about $\sqrt{25}=5$ and to a reduction of the natural $\chi_1$ abundance by about $g_*(T_{\rm FO}^1, \epsilon_1 = 0.001) / g_*(T_{\rm FO}^1, \epsilon_1 = 0.2) \sim 5$. Even though this will increase the lower bound on $M_1$ it could still lead to a better situation in total. To take into account this effect, we will always consider that the entropy producing decays happen very late, i.e., when $g_* \approx 10$.} and hence
\begin{equation}
 \Gamma_{\rm req} \simeq 4.1\cdot 10^{-26} {\rm GeV} \sqrt{g_*} \left( \frac{M_2}{\rm GeV} \right)^2 \left( \frac{\rm keV}{M_1} \right)^2.
 \label{eq:Gamma_req-here}
\end{equation}
There is an important bound on the decay width $\Gamma_2$ to be taken into account: in order not to disturb Big Bang Nucleosynthesis (BBN), the lifetime of the $\chi_2$ decay should be small enough, i.e., it should be smaller than something like 1~sec~\cite{Bezrukov:2009th}. More precisely, the reheating temperature,
\begin{equation}
 T_r\simeq \frac{1}{2} \left( \frac{2 \pi^2 g_*}{45} \right)^{-1/4} \sqrt{\Gamma_2 M_P},
 \label{eq:T_r}
\end{equation}
due to $\chi_2$ decays should be larger than about 0.7~MeV~\cite{Kawasaki:2000en} (weak BBN bound) to 4.0~MeV~\cite{Hannestad:2004px} (strong BBN bound), depending on what is taken into account in the analysis. If this is the case, then the Universe simply continues to cool down after the entropy production has been finished, and will at some point undergo BBN in the usual way.\footnote{Note that there is one additional relativistic degree of freedom during BBN, namely the keVin $\chi_1$. However, since the abundance of this particle gets diluted by the entropy production, this is not going to have a big influence on BBN.} In our case, this translates into a minimal value of the decay width of
\begin{equation}
 \Gamma_{\rm min} = \sqrt{g_*} (3.5\cdot 10^{-24}\ {\rm GeV}, 1.1\cdot 10^{-25}\ {\rm GeV}),
 \label{eq:Gamma_min}
\end{equation}
for $T_r = (4.0, 0.7)$~MeV. One can see from Eqs.~\eqref{eq:Gamma_req-here} and~\eqref{eq:Gamma_min} that the ``natural'' value of $\Gamma_{\rm req}$ (for $M_1=1$~keV and $M_2=1$~GeV) is actually \emph{below} the bound $\Gamma_{\rm min}$. However, this is not a problem as long as we can choose the masses $M_1$ and $M_2$ such that we are okay with all bounds.

\begin{figure}[t]
\centering
\includegraphics[width=7.8cm]{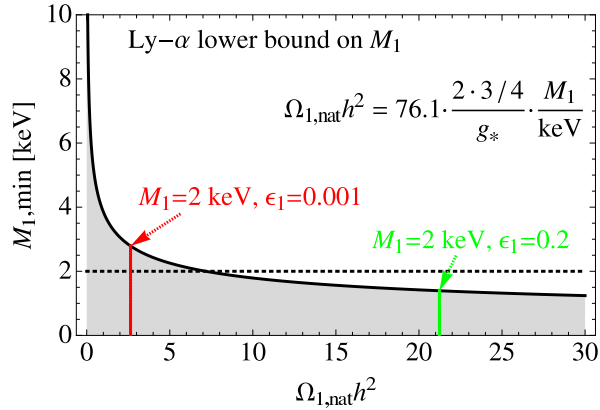}
\caption{\label{fig:Lya} The Ly--$\alpha$ lower bound on $M_1$, as function of the natural abundance for certain parameter combinations. A working (green) and an excluded (red) parameter combination are indicated.}
\end{figure}

Requiring $\Gamma_{\rm req} > \Gamma_{\rm min}$, one obtains the relation
\begin{equation}
 \frac{M_2}{\rm GeV} > (9.2, 1.6)\ \frac{M_1}{\rm keV}.
 \label{eq:M12-rel}
\end{equation}
Successful structure formation (i.e.\ the correct velocity distribution of the DM particles) is probed by the so-called Lyman--$\alpha$ (Ly--$\alpha$) bound~\cite{Boyarsky:2008xj}. As explained in Ref.~\cite{Bezrukov:2009th}, we have to rescale the bound of 8~keV obtained in Ref.~\cite{Boyarsky:2008xj} for non-resonantly produced WDM particles by a factor $\mathcal{S}_{\rm req}^{-1/3}$, where $\mathcal{S}_{\rm req}$ is the entropy dilution factor required to meet the correct DM abundance. The corresponding lower bound on $M_1$ is displayed in Fig.~\ref{fig:Lya}, where we have used the two example values $\epsilon_1 = 0.001$ and $\epsilon_1 = 0.2$, illustrating the robustness of the bound. Since the latter value is very close to the upper bound on $\epsilon_1$ from the $Z$--boson decay width, we can conclude that the lowest possible value for $M_1$ is around $1.5$~keV, which is practically identical to the bound obtained in Ref.~\cite{Bezrukov:2009th}. Hence, an absolute lower bound on $M_2$ can be obtained:
\begin{equation}
 M_2 > (13.8~{\rm GeV} , 2.4~{\rm GeV} ).
 \label{eq:M2_bound}
\end{equation}
Glancing at Fig.~\ref{fig:FO-regions}, one can see that the more restrictive (strong) bound pushes the allowed region of $M_2$ and $\epsilon_2$ into a very small region in the upper left corner of the relativistic freeze-out region. This does, however, not mean that the whole mechanism does not work, but it means that we would need to do a full numerical analysis of the Boltzmann equation to treat the intermediate freeze-out region properly. On the other hand, the less restrictive (weak) bound leaves us with a much larger region that we can calculate, where $\epsilon_2$ can be as large as approximately $5\cdot 10^{-3}$.

\begin{figure}[t]
\centering
\includegraphics[width=7.8cm]{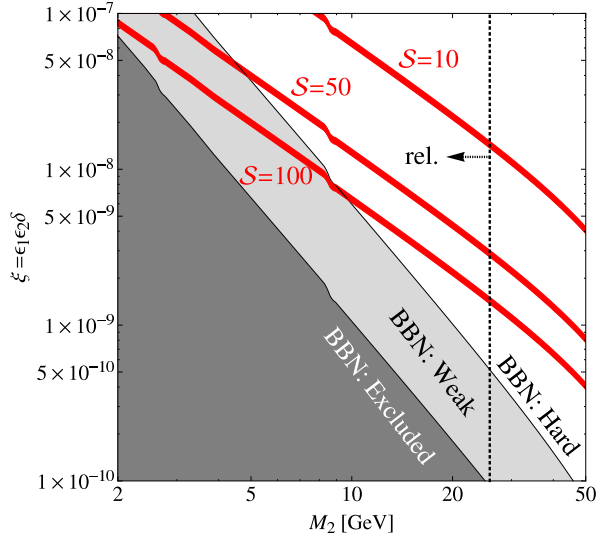}
\caption{\label{fig:En_Regions} Approximate regions where potentially much entropy could be produced, for the case of relativistic freeze-out of $\chi_2$. Note, however, that the vertical line only denotes the value of $M_2$ below which relativistic freeze-out could be possible at all, while the exact bound also depends on the value of $\epsilon_2$ (cf.\ Fig.~\ref{fig:FO-regions}).}
\end{figure}

To give an estimate for our parameters, we can use the low-energy approximation of Eq.~\eqref{eq:f-decay} as estimate for the full decay width,
\begin{equation}
 \Gamma_2 \approx \frac{g^4 \xi^2 \Sigma M_2^5}{1536 \pi^3 c_W^2 M_Z^4} \simeq 1.1\cdot 10^{-12}~{\rm GeV}\ \xi^2\ \left(\frac{M_2}{\rm GeV}\right)^5.
 \label{eq:G2_approx}
\end{equation}
Translating this into a value for $\xi = \epsilon_1 \epsilon_2 \delta$, one obtains
\begin{equation}
 \xi \simeq 2.0\cdot 10^{-7} g_*^{1/4} \left( \frac{\rm GeV}{M_2} \right)^{3/2} \frac{\rm keV}{M_1}.
 \label{eq:xi_val}
\end{equation}
Hence, e.g.\ for $M_1=2$~keV and $M_2=14$~GeV, one obtains $\xi = 3.6\cdot 10^{-9}$. $M_1$ is
consistent with the Ly--$\alpha$ bound for $\epsilon_1 = 0.2$, and the value of $M_2$ restricts $\epsilon_2$ to be something like $8\cdot 10^{-5}$, in which case we obtain $\delta = 2.2\cdot 10^{-4}$. We observe that, because the allowed range of $\epsilon_2$ is quite restricted, we need a strong hierarchy between the two suppressions, $\epsilon_1 \ll \epsilon_2$, for relatively large values of $\epsilon_1$.

The estimate for the situation with relativistic freeze-out of $\chi_2$ is also depicted in Fig.~\ref{fig:En_Regions}. Using the analytical estimates from above, one can already see that the allowed regions for entropy production factors of 100 or more are not very big. This tension will even become stronger when taking a closer look: in Fig.~\ref{fig:En_Regions}, the region left of the vertical line is where a relativistic freeze-out of $\chi_2$ \emph{could} be possible. However, if it does indeed happens will also depend on the value of the coupling constant $\epsilon_2$ itself, and we will see that this will in the end make it hard to conform with the hard BBN bound on the reheating temperature. However, this bound can still be avoided in some circumstances~\cite{Hannestad:2004px}, which depends on the details of the reheating process. A dedicated investigation of this aspect is beyond the scope of this paper, but we would like to suggest it as possible further study.

\subsection{\label{sec:keVins_numerical}The numerical analysis}

We will now perform a more detailed analysis of our model. For certain scenarios, i.e.\ certain values for the parameters $(\epsilon_1, \delta)$, we have first calculated the freeze-out temperatures, then determined the natural $\chi_1$ and $\chi_2$ abundances, calculated the possible amount of entropy production (for certain combinations of $M_2$ and $\epsilon_2$), and finally obtained the diluted $\chi_1$ abundance for a certain value of $M_1$. Note that, although $\Omega_{\chi_1} h^2 \propto M_1$ [cf.\ Eq.~\eqref{eq:1-abundance}], $M_1$ is still essentially unconstrained by the other parameter values, as indicated in Fig.~\ref{fig:FO-regions}. Hence, as long as we can avoid the Ly--$\alpha$ bound, cf.\ Fig.~\ref{fig:Lya}, we can more or less select between a range of different choices for $M_1$. This will of course affect the final $\chi_1$ abundance, but it will not affect the amount of entropy production, so that in one and the same scenario, i.e.\ for the same combination $(\epsilon_1, \delta)$, one might be able to hit the observed value of the DM abundance for relativistic freeze-out of $\chi_2$ (small $\epsilon_2$ to have early freeze-out, but larger $M_2$ to have a fast enough $\chi_2$ decay) with larger $M_1$, or for non-relativistic freeze-out of $\chi_2$ (large $\epsilon_2$ to have late freeze-out, but smaller $M_2$ to have a slow enough $\chi_2$ decay) with smaller $M_1$. The lesson to learn is that, although one might naively favour relativistic freeze-out of $\chi_2$ to have a high enough abundance, it turns out that with non-relativistic freeze-out it is much easier to avoid the BBN bound, so depending on the actual choice of parameters one or the other situation could be of advantage.

\begin{figure}[t]
\centering
\includegraphics[width=7.8cm]{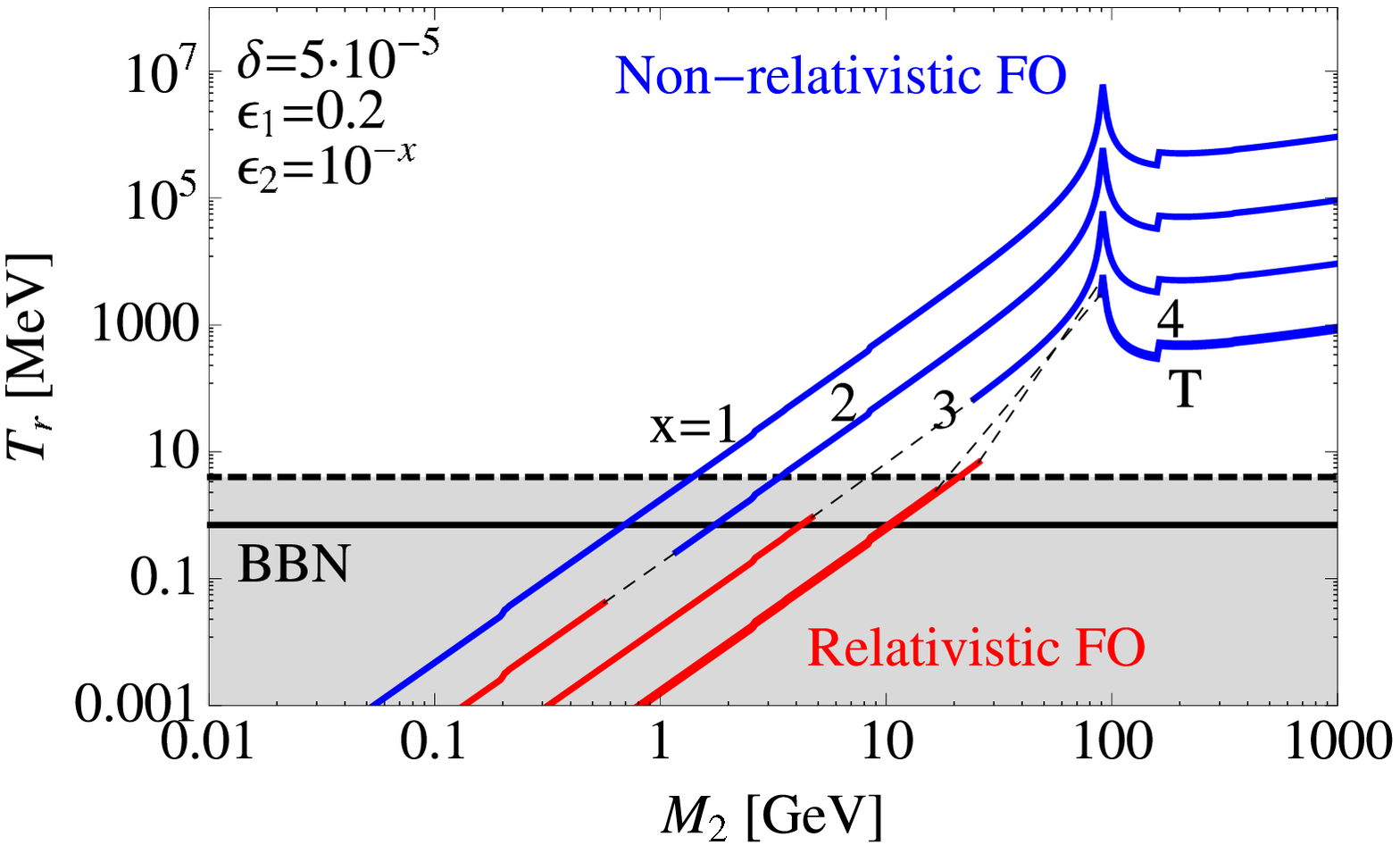}
\includegraphics[width=7.8cm]{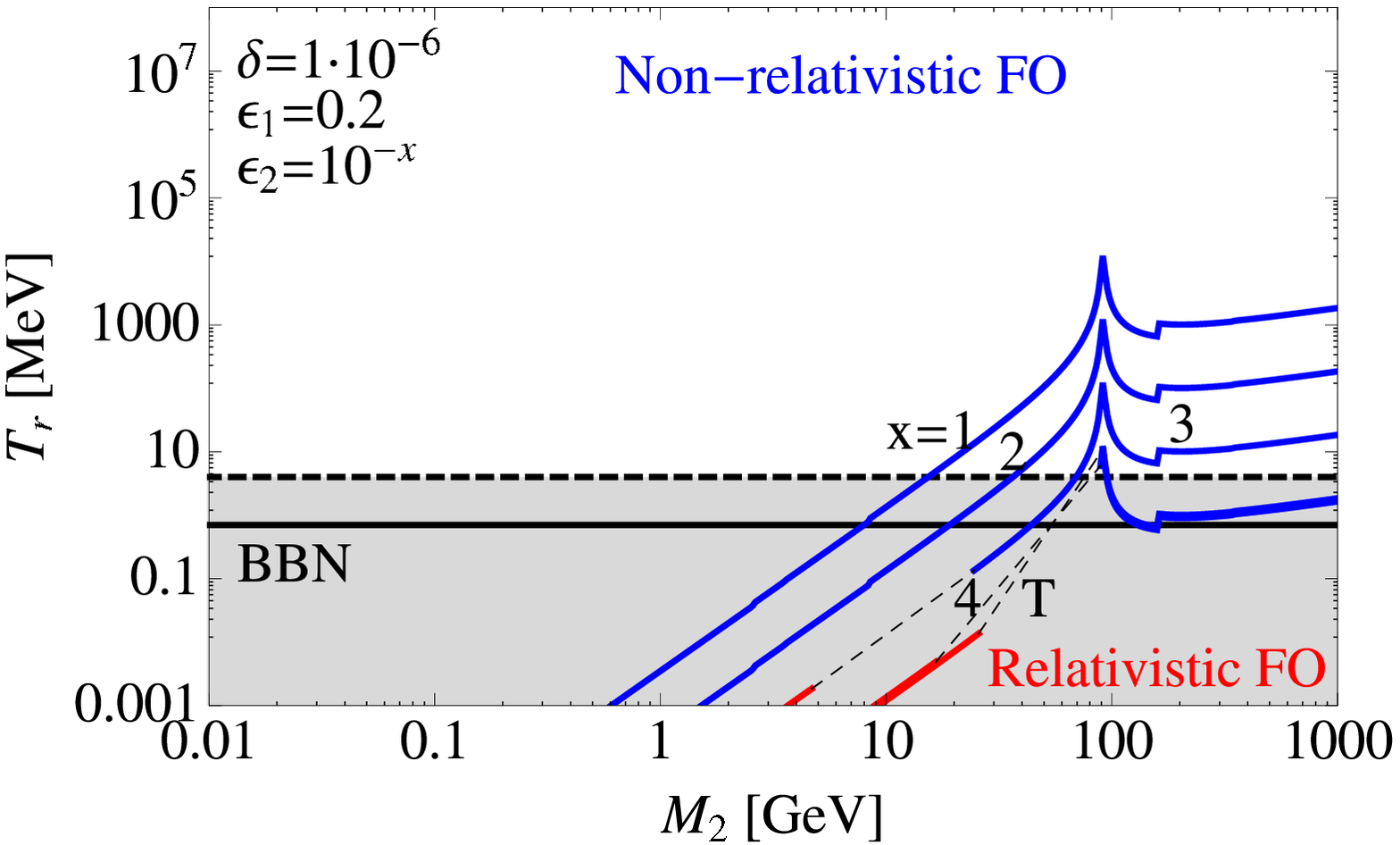}
\includegraphics[width=7.8cm]{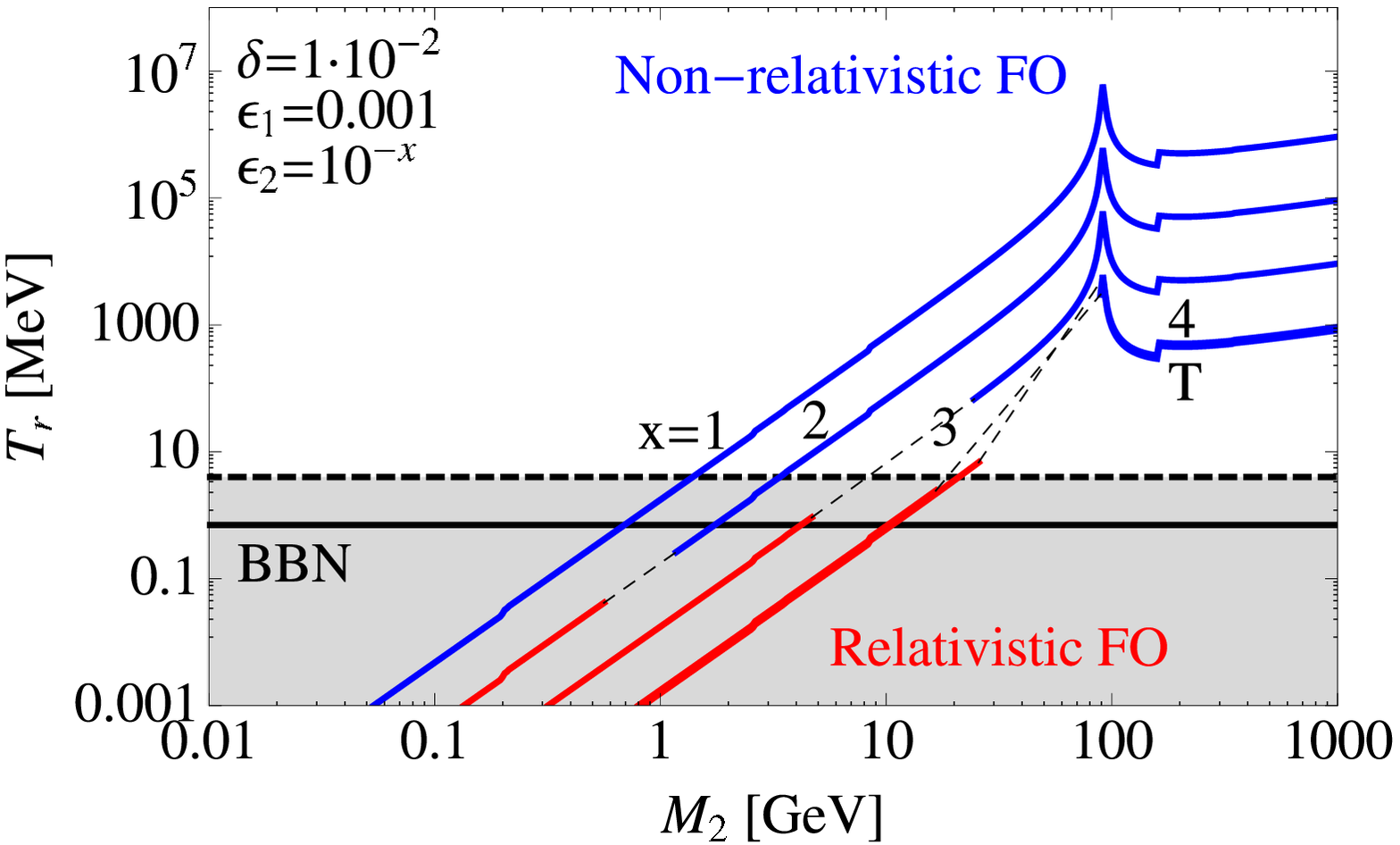}
\includegraphics[width=7.8cm]{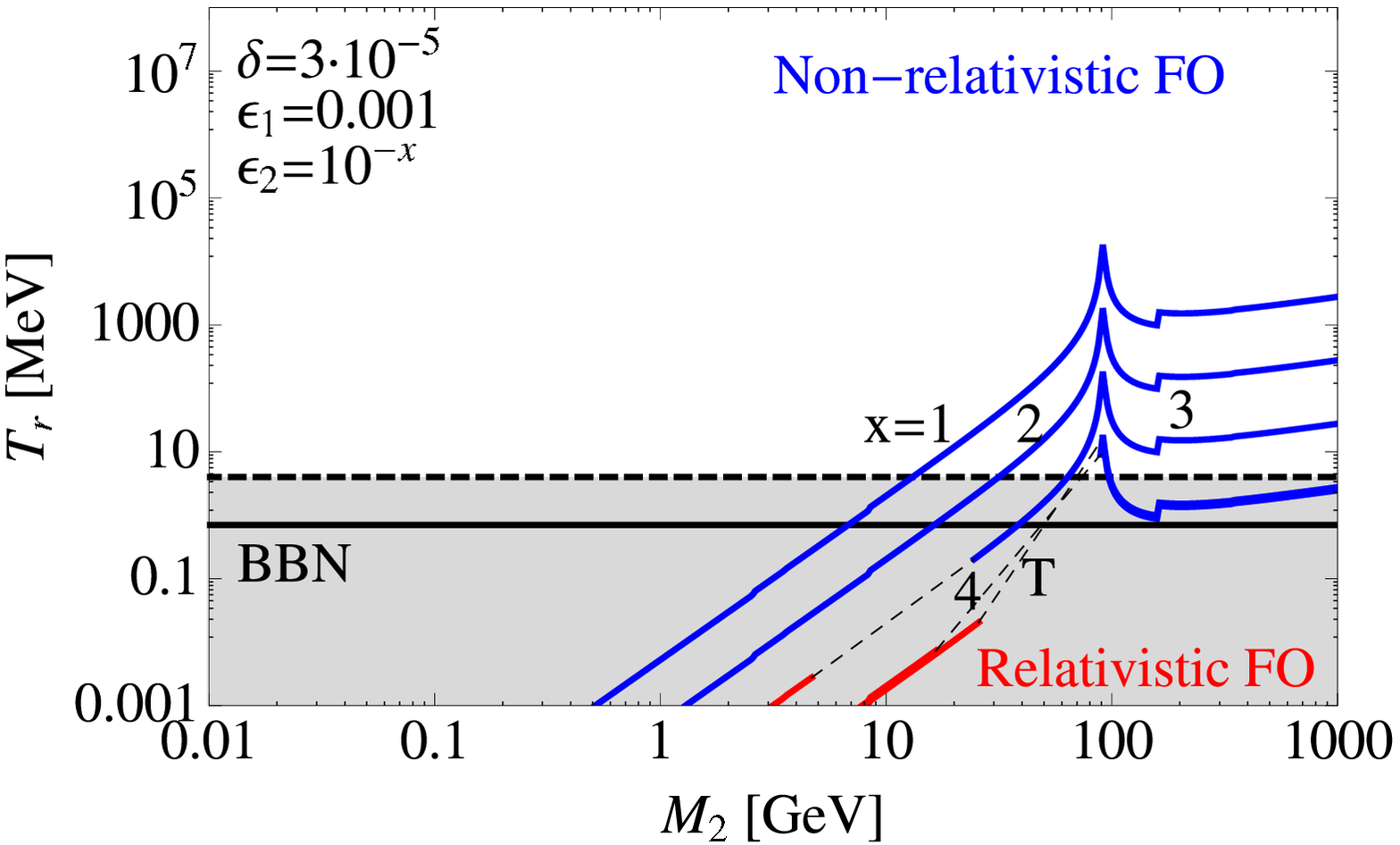}
\caption{\label{fig:TR} Reheating temperatures after $\chi_2$-decay.  The gray patches mark the BBN bound (dashed: hard bound, $T_r > 4.0$~MeV; solid: weak bound, $T_r > 0.7$~MeV). (See text for further explanations.)}
\end{figure}

Let us first have a look at the correponding reheating temperatures, which are depicted in Fig.~\ref{fig:TR}. Here, we have calculated Eq.~\eqref{eq:T_r} for different combinations of the important parameters. The colour code is that red lines signal relativistic (hot) freeze-out of the GeVin $\chi_2$, while blue lines stand for non-relativistic (cold) freeze-out of $\chi_2$. The intermediate freeze-out region (cf.\ light gray band in Fig.~\ref{fig:FO-regions}) has not been calculated by us, due to the lack of an approximate solution of the Boltzmann equation, but it is indicated in the plots by the thin dashed black lines. However, these lines simply interpolate between the hot and cold freeze-out regions, and they can only serve as guide for the eye but do not signal the physical reheating temperatures in that region. As we can see from the plots, it is extremely difficult to have a high enough reheating temperature for relativistic freeze-out of $\chi_2$, while this is relatively easily achieved for non-relativistic freeze-out. However, this does not yet give information about how much entropy can actually be produced in the $\chi_2$ decays, which is what we will have a look at next.

\begin{figure}[t]
\centering
\includegraphics[width=7.8cm]{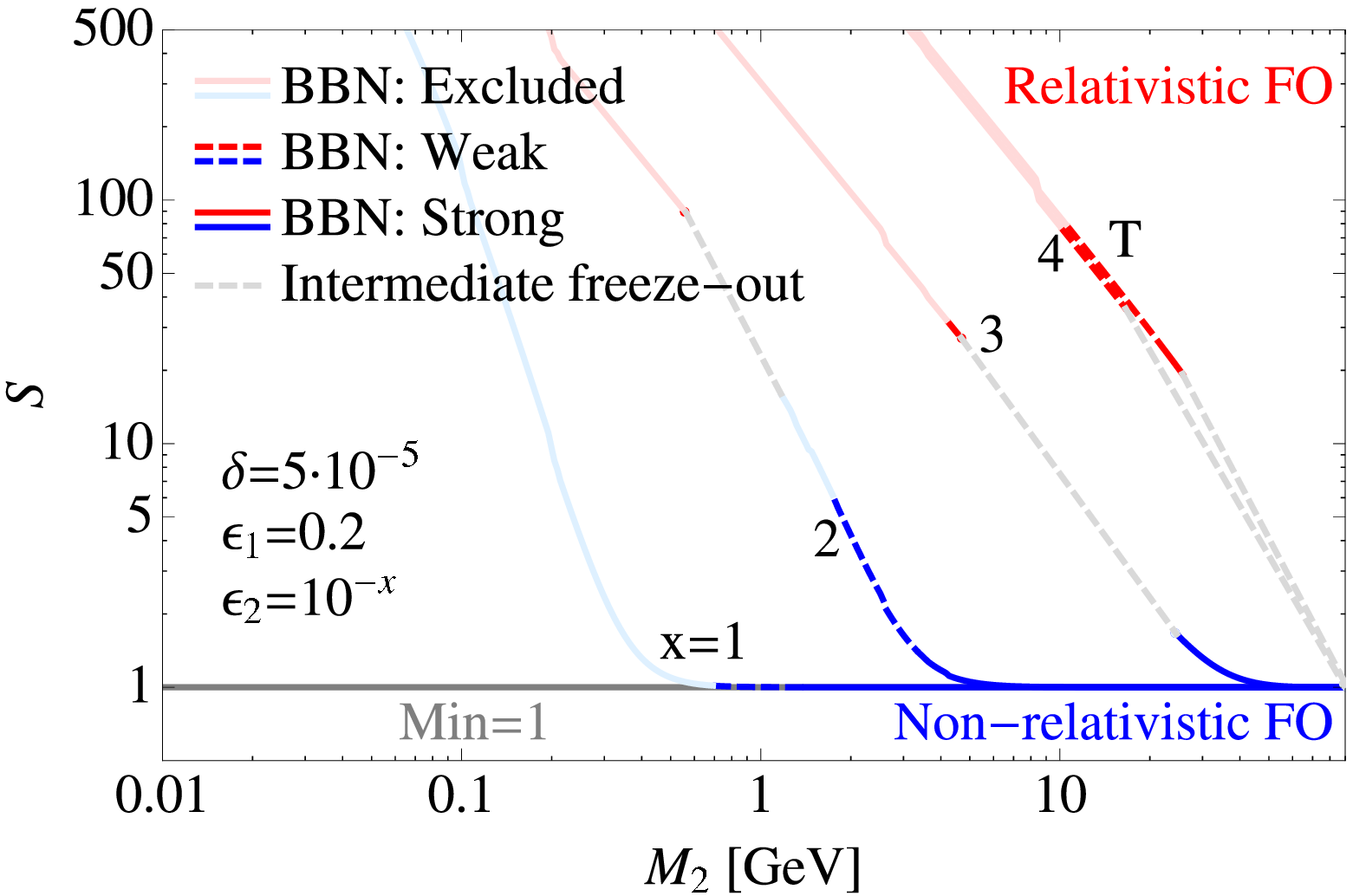}
\includegraphics[width=7.8cm]{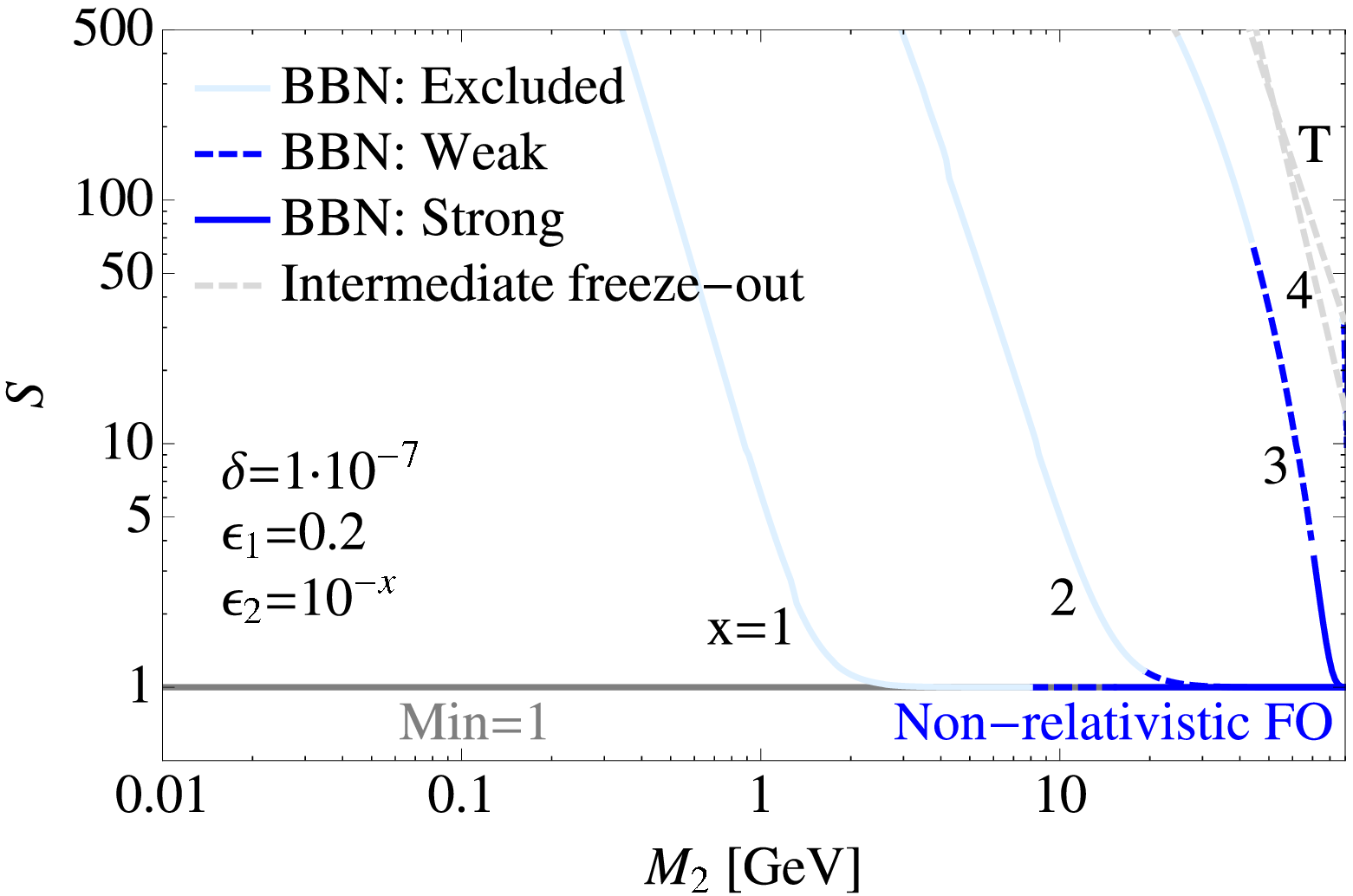}
\includegraphics[width=7.8cm]{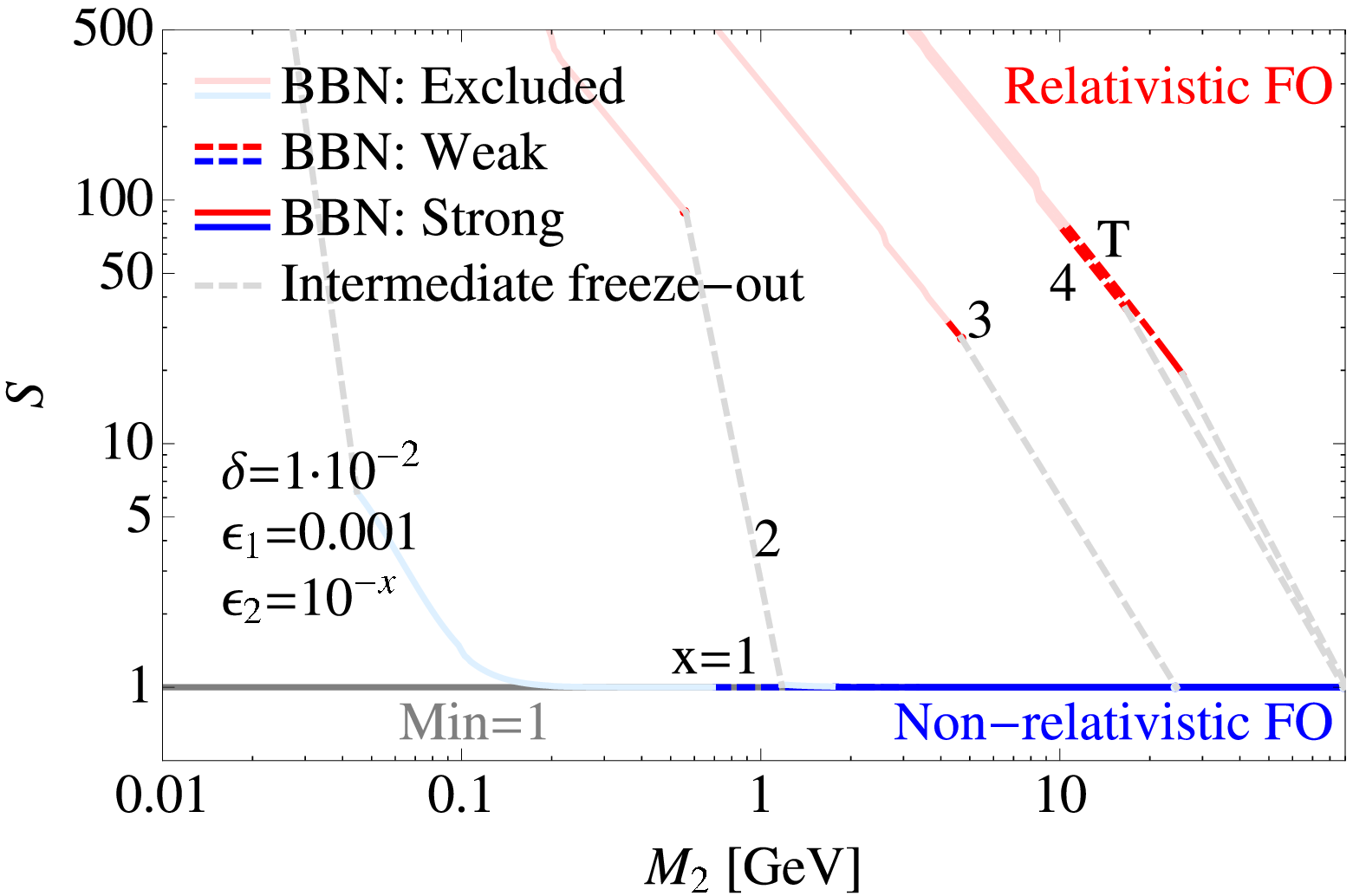}
\includegraphics[width=7.8cm]{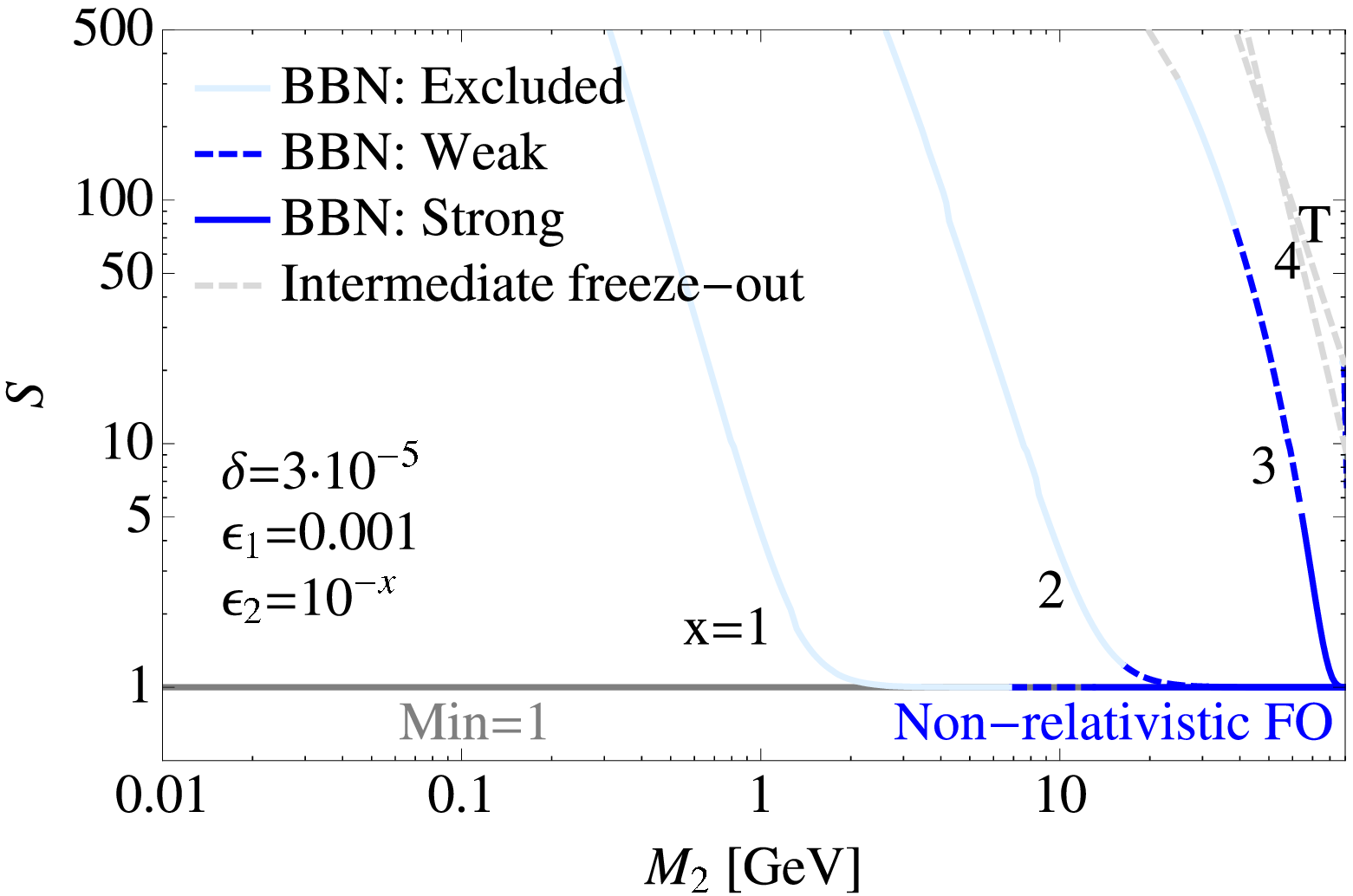}
\caption{\label{fig:Entropy} Entropy production for different values of $(M_2, \epsilon_2)$ in a given scenario, for the cases of relativistic (red) and non-relativistic (blue) freeze-out of $\chi_2$ (the interpolation for the intermediate freeze-out region is indicated by the dashed gray line). Dashed dark lines are consistent with the weak BBN bound, while solid lines are even consistent with the strong bound. Everything printed in light colours is excluded by BBN, i.e., the corresponding regions are inconsistent even with the weak bound.}
\end{figure}

The allowed values for the entropy dilution factor, for the same scenarios as in Fig.~\ref{fig:TR}, are depicted in Fig.~\ref{fig:Entropy}. We have plotted the amount of entropy produced in the $\chi_2$ decays [i.e., the factor $\mathcal{S}$ in Eq.~\eqref{eq:ent-prod}] for different values of $M_2$ and $\epsilon_2$, where $T=4.05$ denotes the thermalization limit leading to the smallest possible value of $\epsilon_2$ (cf.\ Fig.~\ref{fig:FO-regions}). As already explained earlier, we have only calculated the cases of relativistic (red) and non-relativistic (blue) freeze-out of $\chi_2$, whereas the intermediate region is only indicated by the dashed light gray lines (which are again only a guide for the eye interpolating between the two limiting cases). Nevertheless, due to the continuous nature of the functions involved, for cases where there is some allowed region for the entropy production by relativistically frozen-out $\chi_2$-particles, there must also be some significant amount of entropy production in the intermediate region, although the true behaviour may not exactly follow our interpolation. Since we, however, wanted to give a proof of principle first, we leave the more detailed numerical calculation for further studies.

As already explained, the amount of entropy that can be produced is limited by the bound from BBN. In Fig.~\ref{fig:Entropy}, dark coloured dashed lines denote the regions that are consistent with the hard BBN bound ($T_r > 4.0$~MeV), while the solid lines denote the regions consistent with the weak BBN bound  ($T_r > 0.7$~MeV) only. The lines drawn in light red and light blue are excluded by BBN. As a general tendency, we see that it is easier to produce enough entropy for the case of relativistic freeze-out of $\chi_2$ (dark red lines), which is exactly contrary to the situation for the reheating temperature in Fig.~\ref{fig:TR}. This can be understood by the suppression of the number density for the case of cold relics. However, a large mass $M_2$ can partially compensate for that, as the abundance itself is the decisive quantity in Eq.~\eqref{eq:ent-prod}, which is why for some choices of parameters there can also be a considerable entropy production for cold freeze-out of the GeVin (cf.\ right panels of Fig.~\ref{fig:Entropy}). However, in general the strong BBN bound is a very restrictive limit that does not allow for much entropy production, as it forces the decay width $\Gamma_2$ to be large. On the other hand, if one only requires the weak BBN bound to be fulfilled, it is indeed possible to produced entropy correction factors up to about $\mathcal{S} \sim 100$.

In principle, it would be possible to produce considerably more entropy by introducing more ``generations'' of GeVins: if there were $N$ particles with a mass similar to $M_2$ and an interaction strength similar to the one of the GeVin, then the entropy correction factor would be multiplied by exactly that number, $\mathcal{S} \to N\cdot \mathcal{S}$ (if $\mathcal{S} \gg 1$). This would not influence the reheating temperature, as it only depends on the value of $\Gamma_2$, cf.\ Eq.~\eqref{eq:T_r}, and it would even strengthen our assumption of the GeVin temporarily dominating the energy density of the Universe. Hence, introducing more types of GeVins could be a natural way to increase the amount of entropy produced considerably. In particular, one could achieve the correct DM abundance while being in accordance even with the hard BBN bound. However, although this is a tempting possibility, we will stick to the minimal situation in order to present an analysis of the simplest framework.

\begin{figure}[t]
\centering
\includegraphics[width=7.8cm]{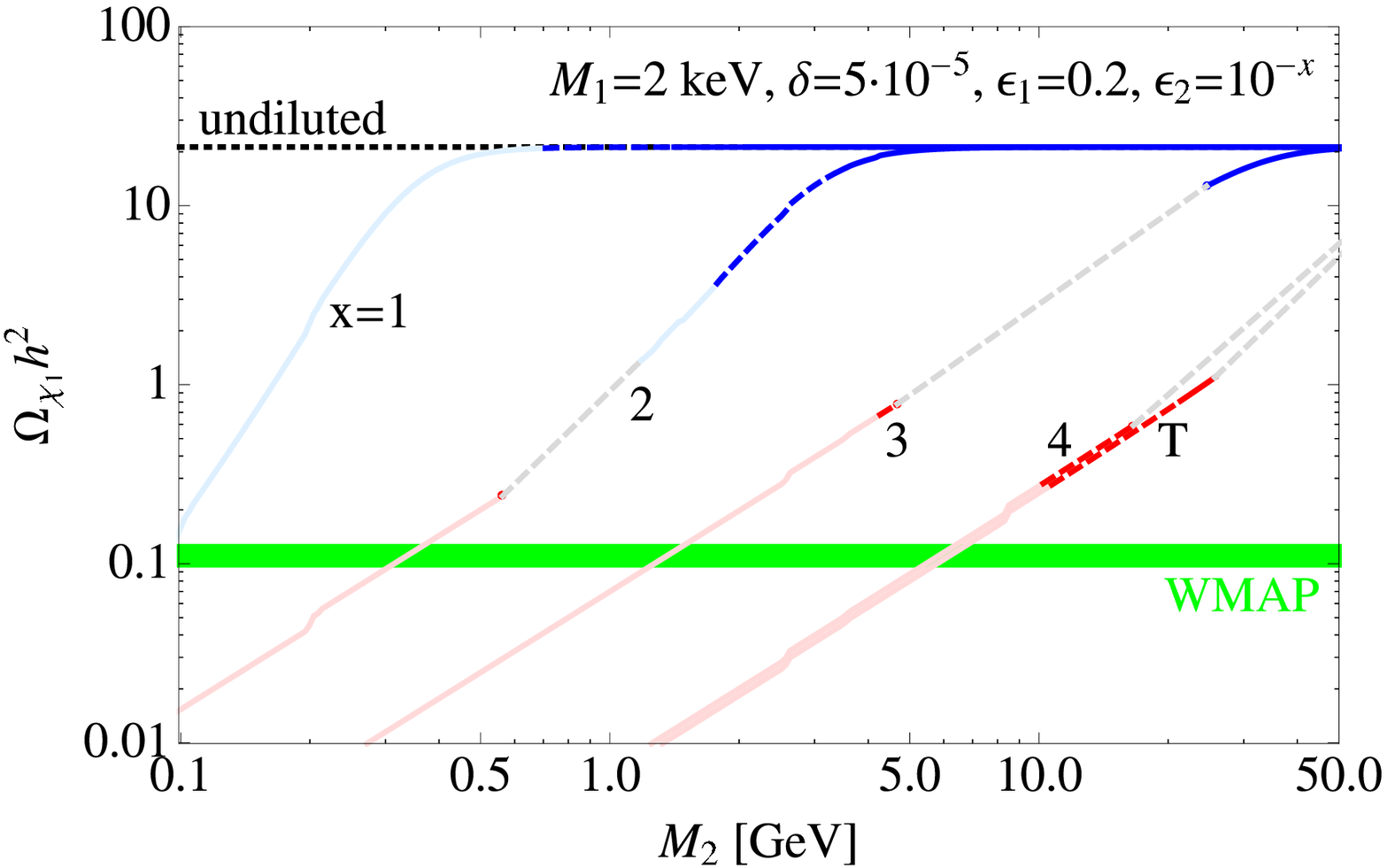}
\includegraphics[width=7.8cm]{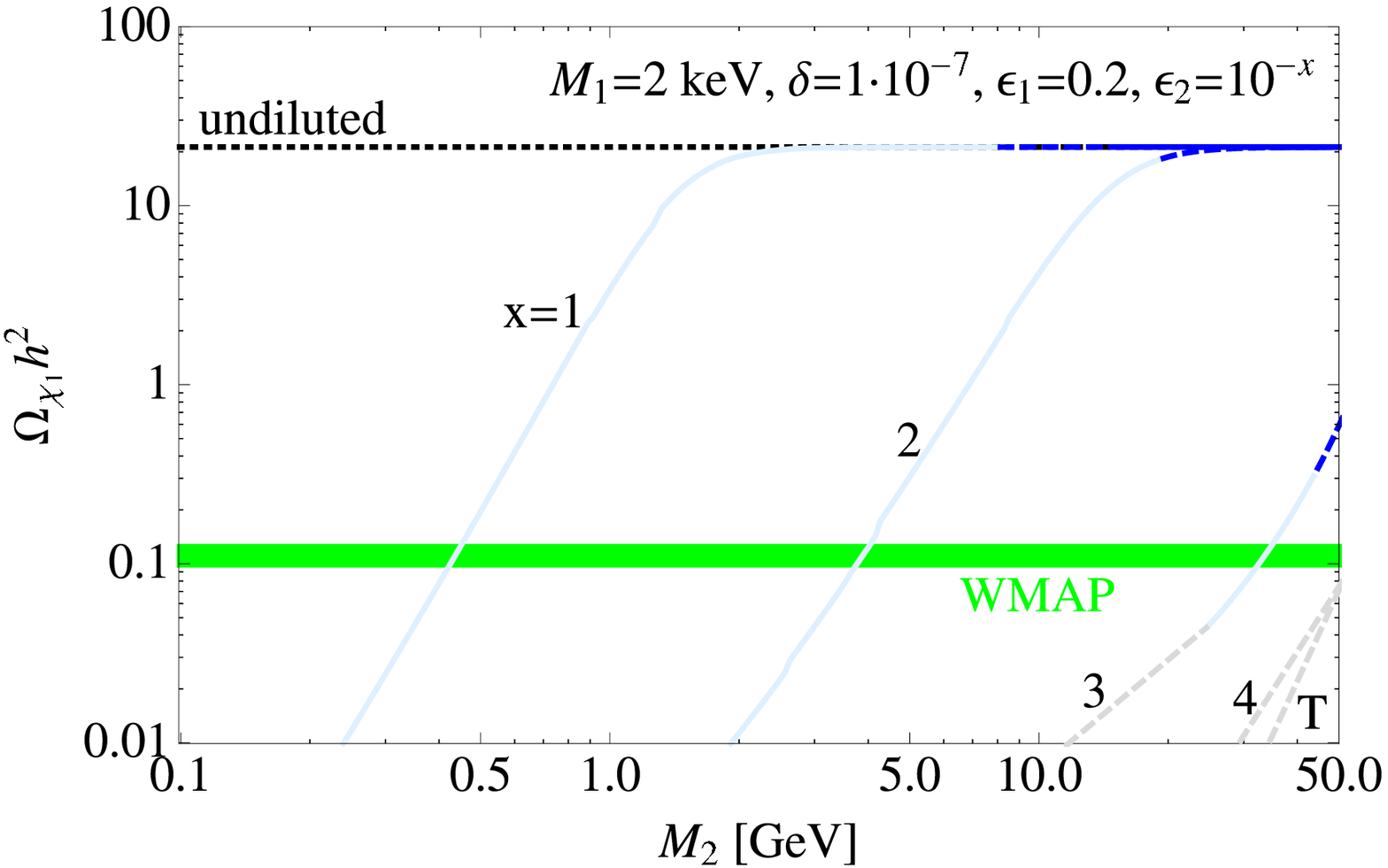}
\includegraphics[width=7.8cm]{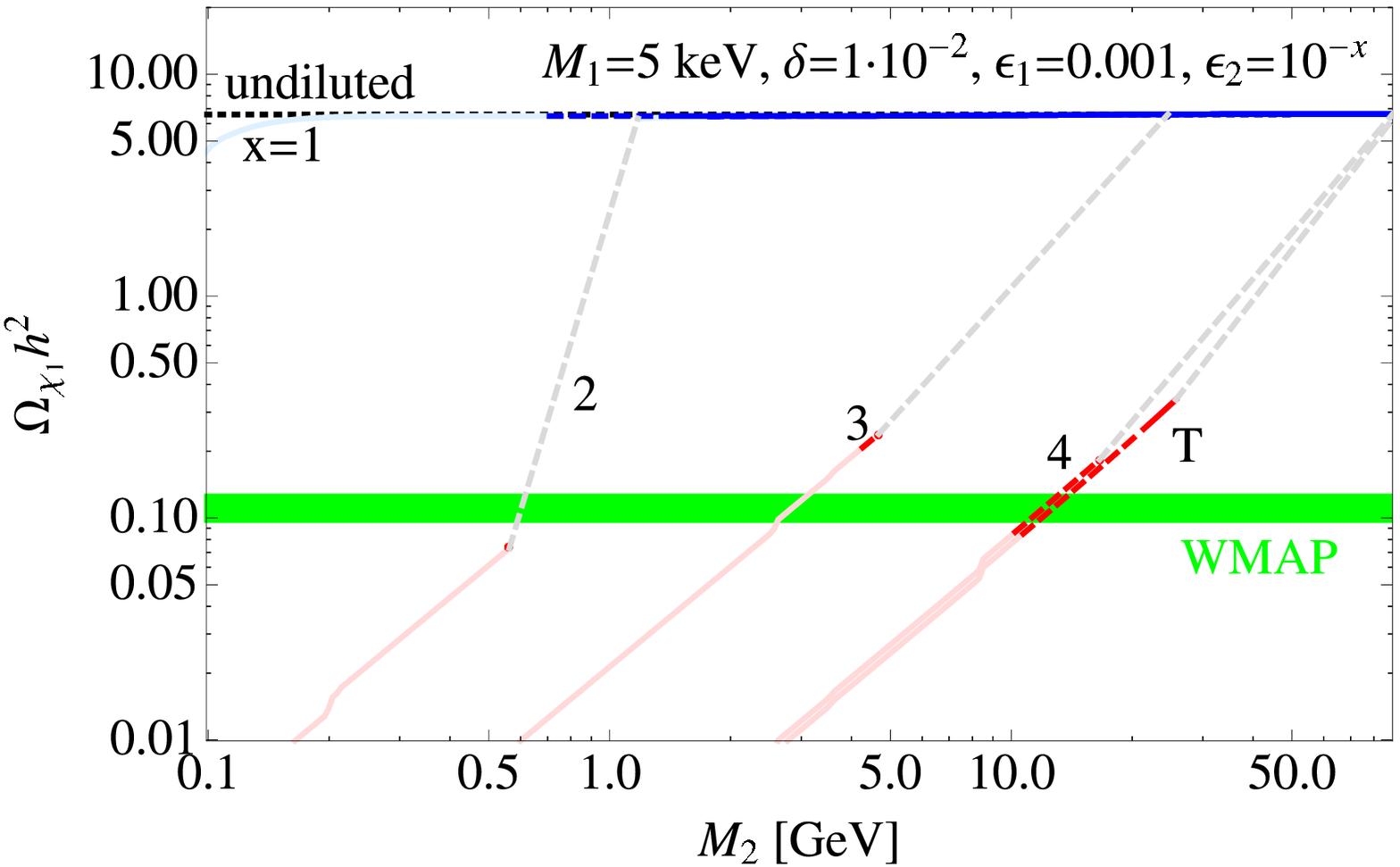}
\includegraphics[width=7.8cm]{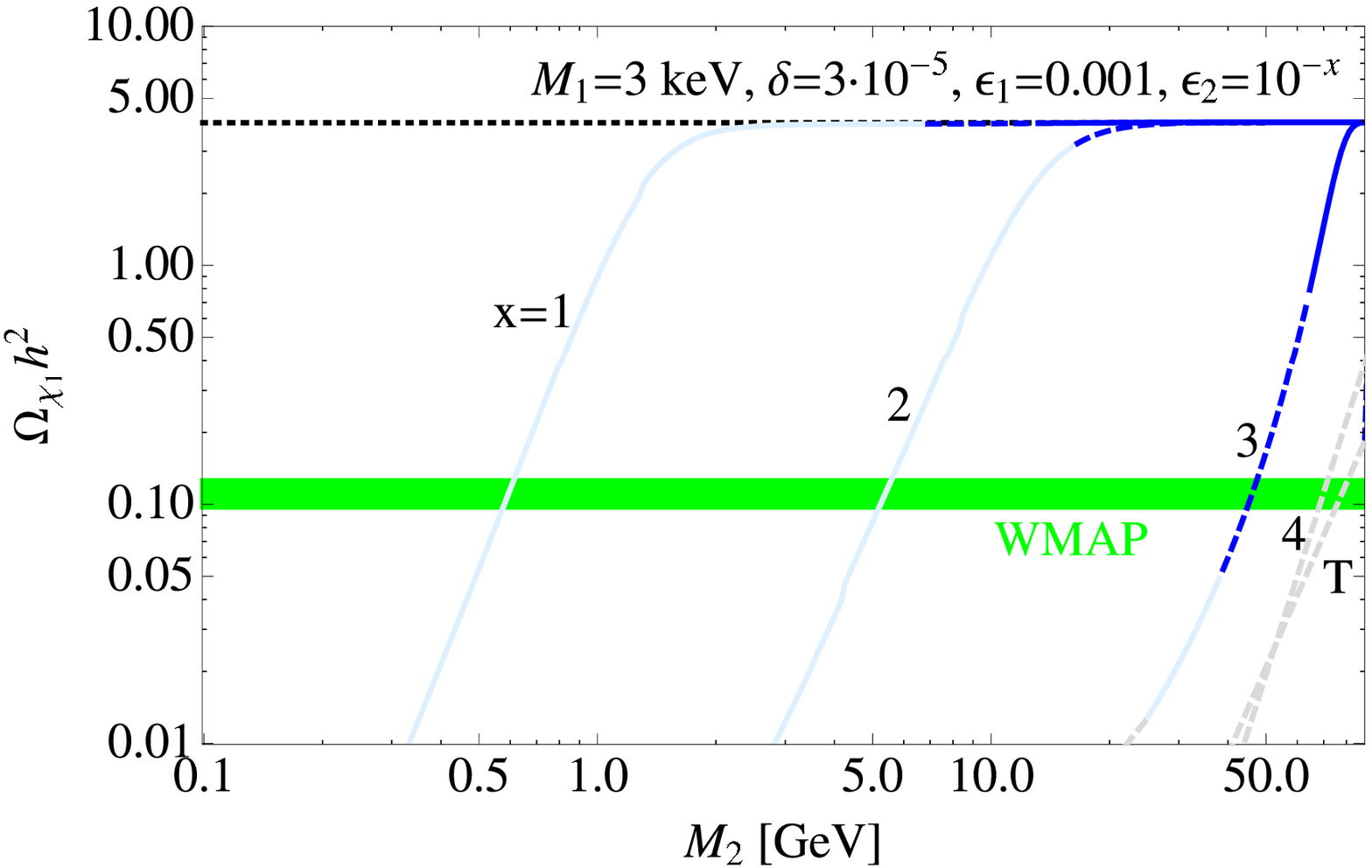}
\caption{\label{fig:Abun} Diluted $\chi_1$ abundance for different values of $(M_2, \epsilon_2)$ in a given scenario, for the cases of relativistic (red) and non-relativistic (blue) freeze-out of $\chi_2$. The natural (undiluted) abundance is marked by the horizontal black dashed line, and in general solid lines mark consistency with both BBN bounds, dashed lines mark consistency with the weak BBN bound only. Light colours signal that the corresponding parameter region is inconsistent with BBN. Gray dashed lines correspond to the intermediate freeze-out region. (See text for further explanations.)}
\end{figure}

The results for the diluted abundance, Eq.~\eqref{eq:ent-prod}, are plotted in Fig.~\ref{fig:Abun}. Again, we have used the color code of red lines signaling relativistic freeze-out of $\chi_2$ while blue lines signal non-relativistic freeze-out of $\chi_2$. We further distinguish between parts of the parameter space that are okay with the strong BBN bound (dark solid lines), parts that are only consistent with the weak bound only (dark dashed lines), and regions that are inconsistent with both of them (light colored lines). The exception to this is the intermediate freeze-out region, which we did not investigate (again, the corresponding gray dashed lines are merely interpolations between the hot and cold regions and serve as guide for the eye only): in case that already the non-relativistic freeze-out region is inconsistent with both BBN bounds, we can be sure that also the intermediate region will not cure that. However, for the other cases we cannot be sure if they agree with the BBN bound, or not. To investigate those regions in detail we refer to a dedicated numerical study of the Boltzmann equation.

Apart from that, we can clearly see from the plots what the effect of the entropy production is: in case that the factor $\mathcal{S}$ is significantly larger than $1$, the abundance of the keVin $\chi_1$ is diluted by the entropy production, and hence decreased compare to its natural value represented by the horizontal black dashed line in Fig.~\ref{fig:Abun}. Depending on the exact values of $M_2$ and $\epsilon_2$, this dilution can be strong enough for the actual $\chi_1$ abundance to meet the $3\sigma$ range allowed by WMAP-7~\cite{Komatsu:2010fb} (green horizontal band in the plots). Note that, although one would naturally expect the relativistic freeze-out of $\chi_2$ to be the favoured situation (as in the left panels of Fig.~\ref{fig:Abun} for values of $\epsilon_2$ close to the thermalization limit), a suppression of the $\chi_2$ decay by a small parameter $\delta$ can be easily cured by a slightly larger value of the mass $M_2$, thereby leading to a more advantageous relation between the decay width $\Gamma_2$ and the energy density $Y_{2,\infty} M_2$ in Eq.~\eqref{eq:ent-prod}. This results into efficient entropy production even for non-relativistic freeze-out of $\chi_2$ (cf.\ lower right panel of Fig.~\ref{fig:Abun}). Note that a further dilution of the natural abundance of the keVins $\chi_1$ can come from a large $g_*$ in Eq.~\eqref{eq:1-abundance}: for very early freeze-out of $\chi_1$, there are more relativistic degrees of freedom available in the Universe, and all of them will be produced. By this, the energy density present is distributed among more relativistic species, leaving less left for $\chi_1$. Finally, as explained before, we have chosen the values of $M_1$ in order to be consistent with the Ly--$\alpha$ bound, but since they are otherwise essentially unconstrained one could also choose different values.

Summing up, one has many possibilities to indeed hit the correct abundance in our setting, which leads us to the conclusion that it would without any doubt be worth to do a dedicated numerical study in order to investigate different scenarios in greater detail. We also plan to perform such a study ourselves in the foreseeable future.

\subsection{\label{sec:keVins_benchmark}Benchmark points}

Let us look in more detail into the promising parameter regions. We have seen from Fig.~\ref{fig:Abun} that in general smaller values of $\epsilon_1$ seem to be preferred. This was to be expected, since smaller $\epsilon_1$ leads to an earlier freeze-out of $\chi_1$ and hence to a larger value of $g_*$, which decreases the $\chi_1$ abundance according to Eq.~\eqref{eq:1-abundance}. Since it is non-trivial to produce large amounts of entropy dilution, i.e.\ $\mathcal{S} \gg 1$, a naturally small abundance of $\chi_1$ is desired. Furthermore, it is of advantage to ``tune'' the decay width $\Gamma_2$ separately by adjusting the parameter $\delta$, in order to keep the reheating temperature high enough while staying close to the lower bound in order to produce enough entropy. Still, from the lower panels of Fig.~\ref{fig:Abun}, we can see that it is possible to hit the correct abundance for both, hot and cold $\chi_2$, and in accordance with the weak BBN bound.

\begin{figure}[t]
\centering
\includegraphics[width=7.8cm]{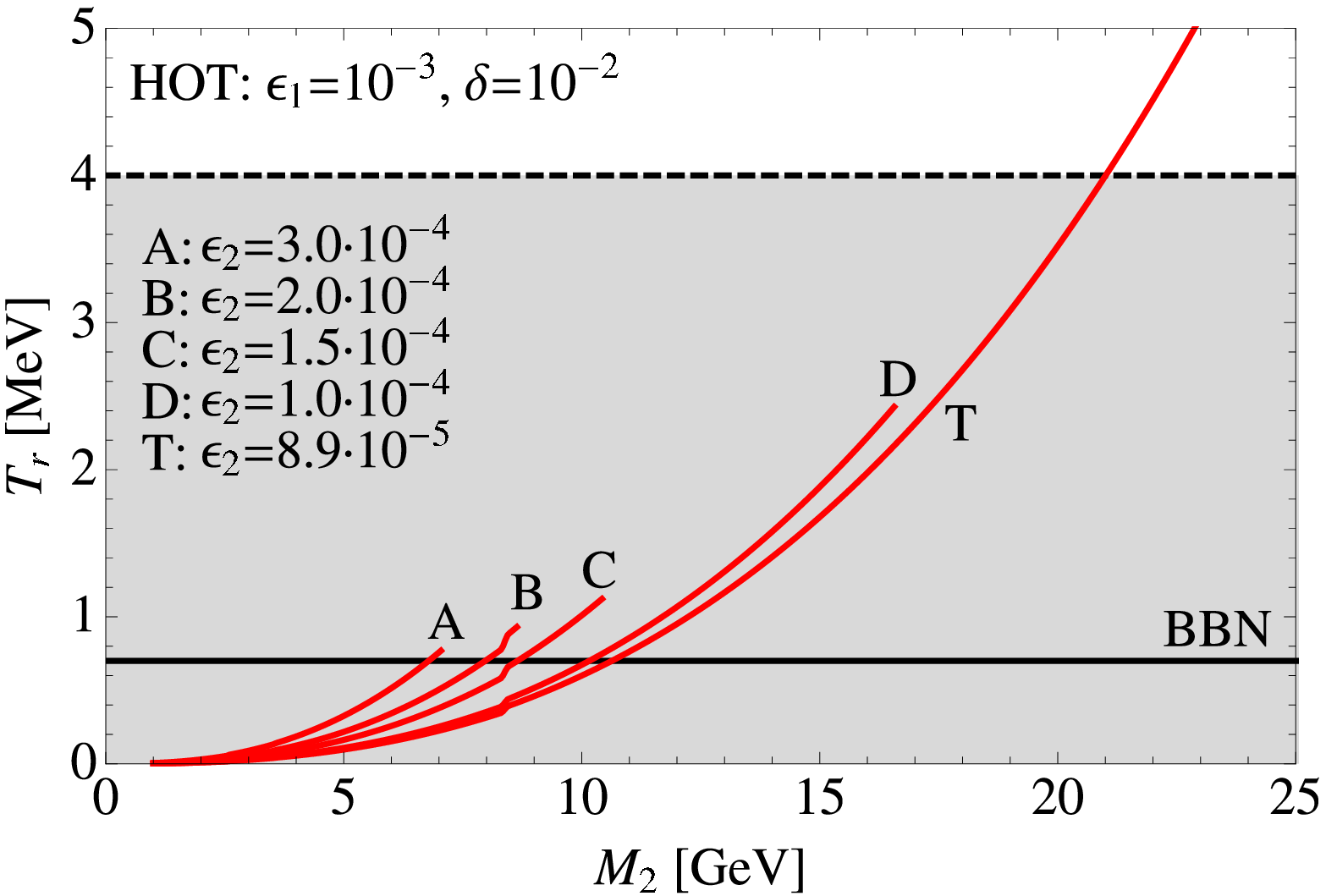}
\includegraphics[width=7.8cm]{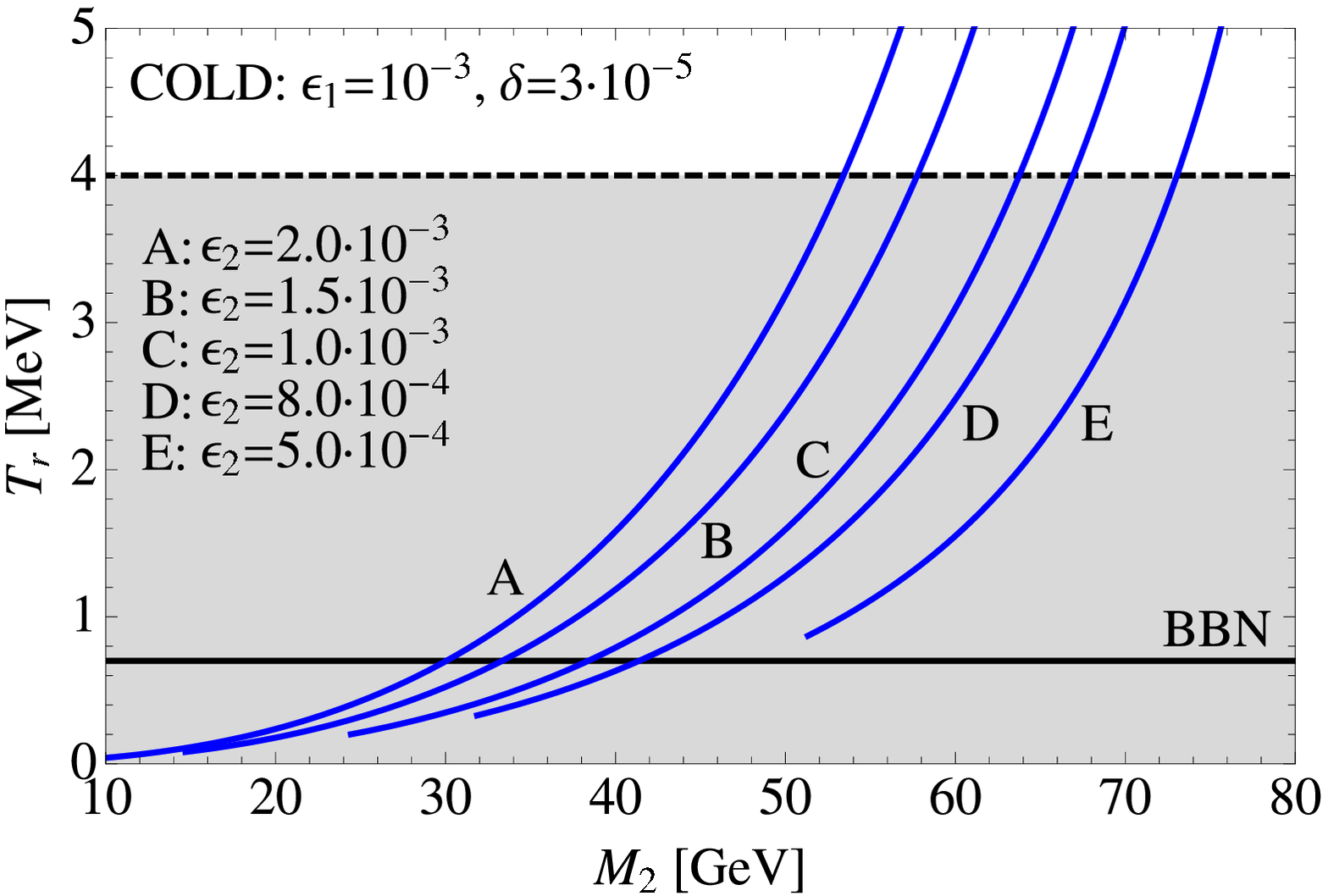}
\includegraphics[width=7.8cm]{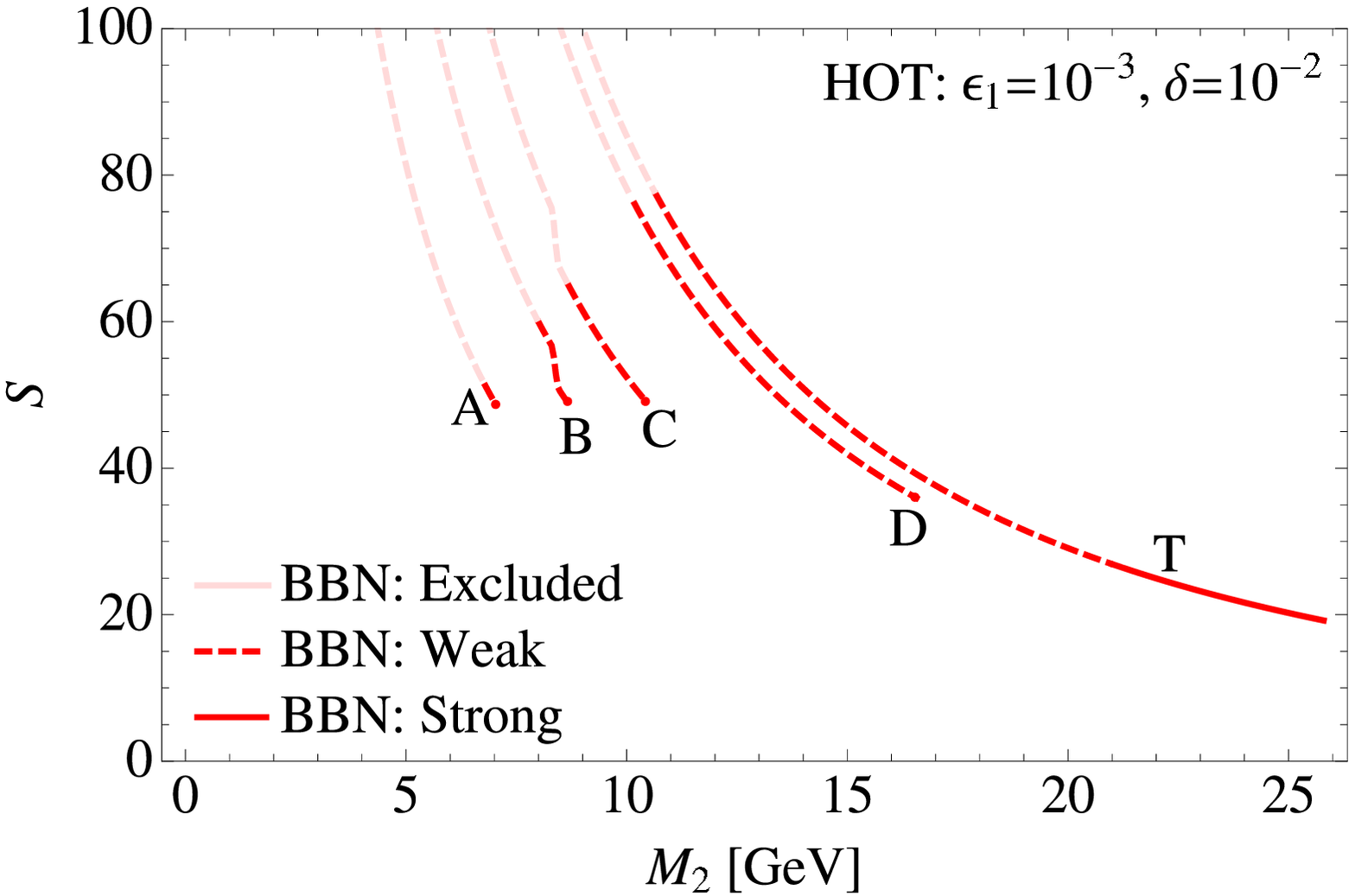}
\includegraphics[width=7.8cm]{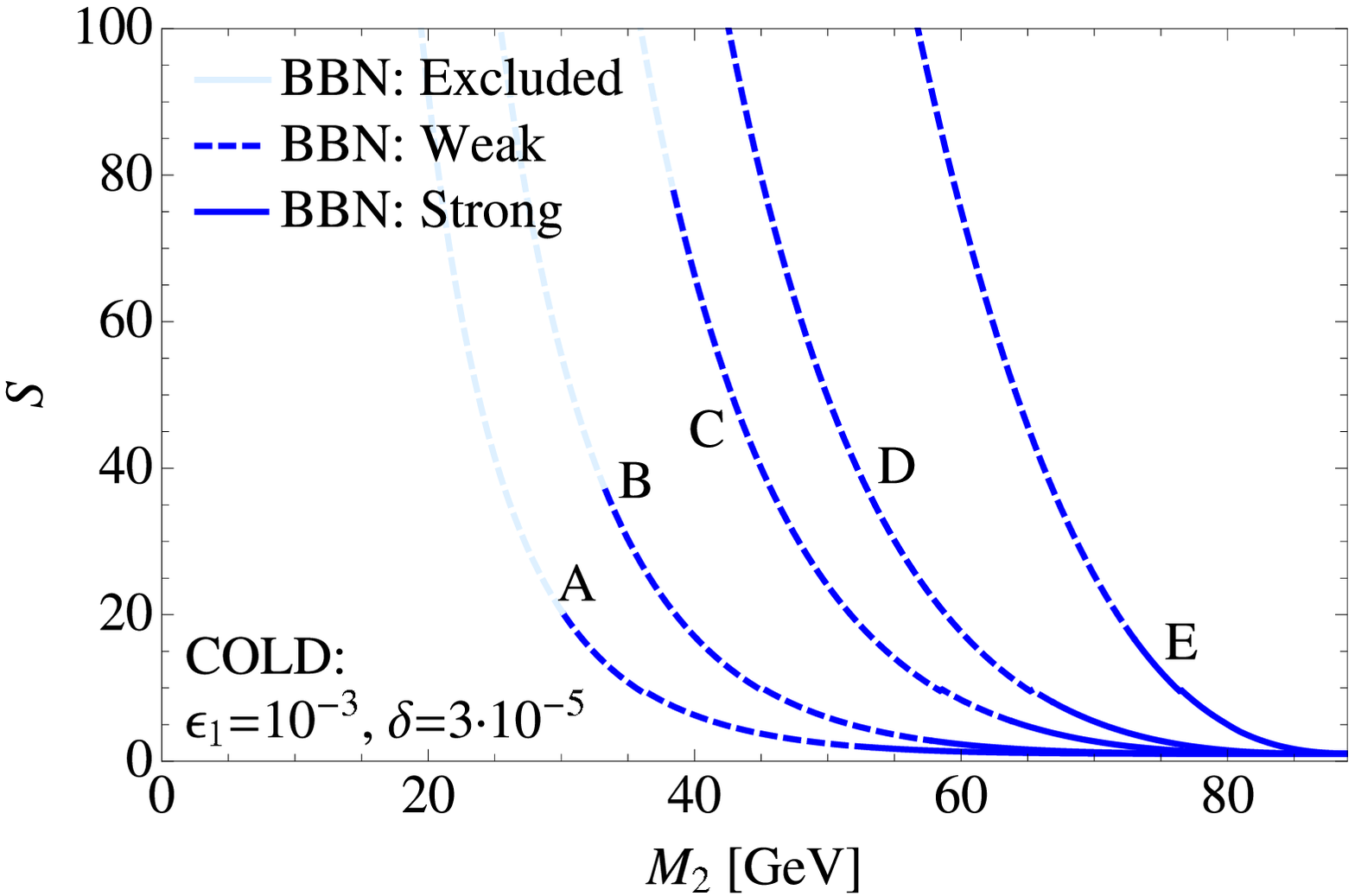}
\caption{\label{fig:Ent_Scen} Freeze-out temperatures and entropy production for the HOT and COLD scenarios. The colour code is the same as in Figs.~\ref{fig:TR} and~\ref{fig:Entropy}.}
\end{figure}

This can be seen more clearly in Fig.~\ref{fig:Ent_Scen}, where the possible freeze-out temperatures (upper panels) and the possible amounts of entropy dilution (lower panels) are depicted, for both the HOT (left panels) and the COLD (right panels) scenarios, defined in the figure, which correspond to relativistic and non-relativistic freeze-out of $\chi_2$, respectively. This time, we have decided for a linear scale in order to illustrate that the region in parameter space leading to significant entropy production is not infinitely small. Note that in particular for the HOT scenario, it might very well be possible to find patches in the parameter space corresponding to the intermediate (warm) freeze-out region of $\chi_2$, which could still be in agreement with all bounds and lead to a significant entropy production. Overall one observes that it is difficult to simultaneously produce large amounts of entropy while still being in agreement with the hard BBN bound, in accordance with our findings from Sec.~\ref{sec:keVins_analytical}. However, the weak BBN bound can easily be fulfilled, and in any case a very detailed derivation of this bound, preferably within the framework of our model, is currently not available. This offers a great possibility to falsify our considerations.

\begin{figure}[t]
\centering
\includegraphics[width=7.8cm]{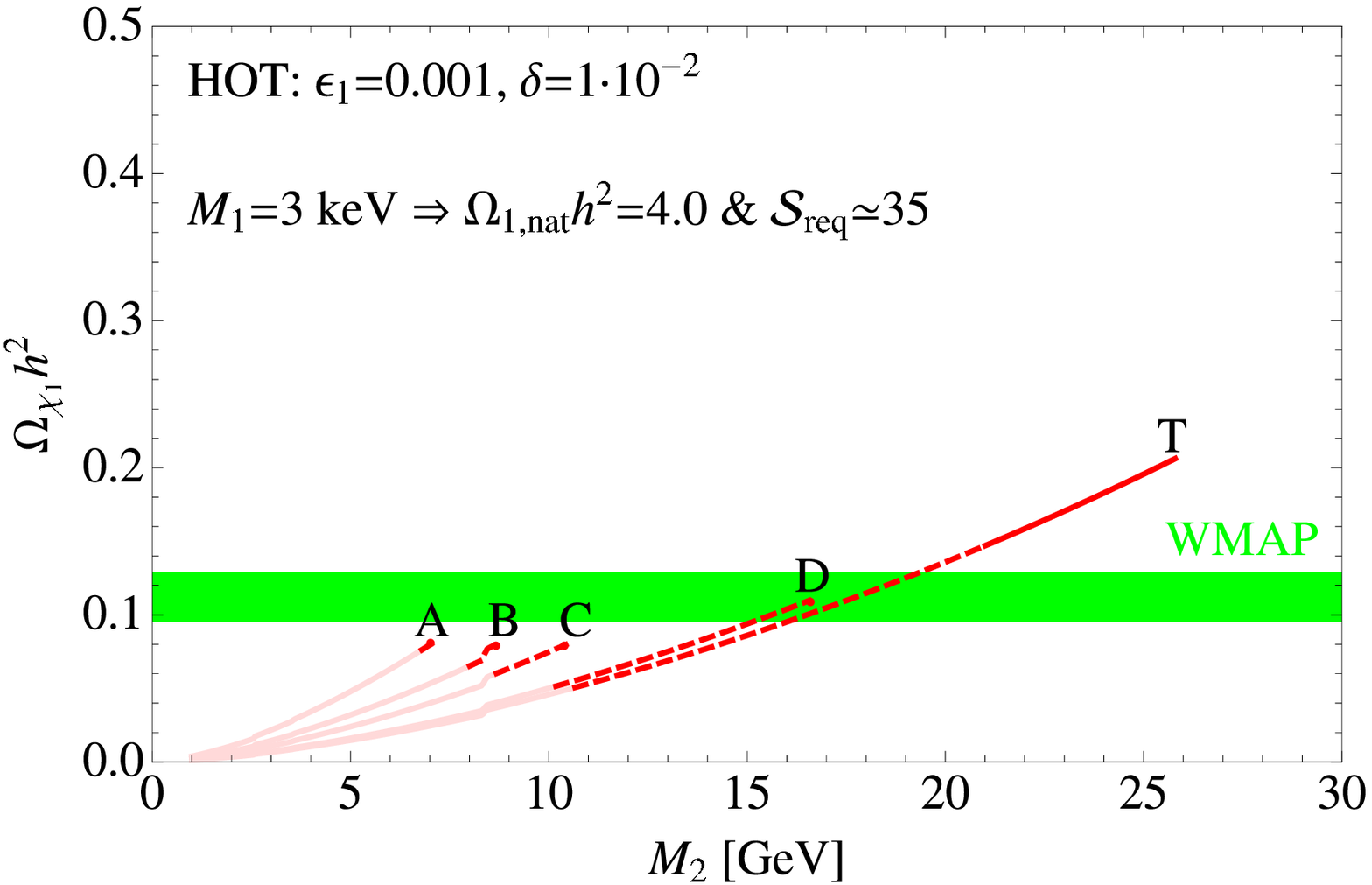}
\includegraphics[width=7.8cm]{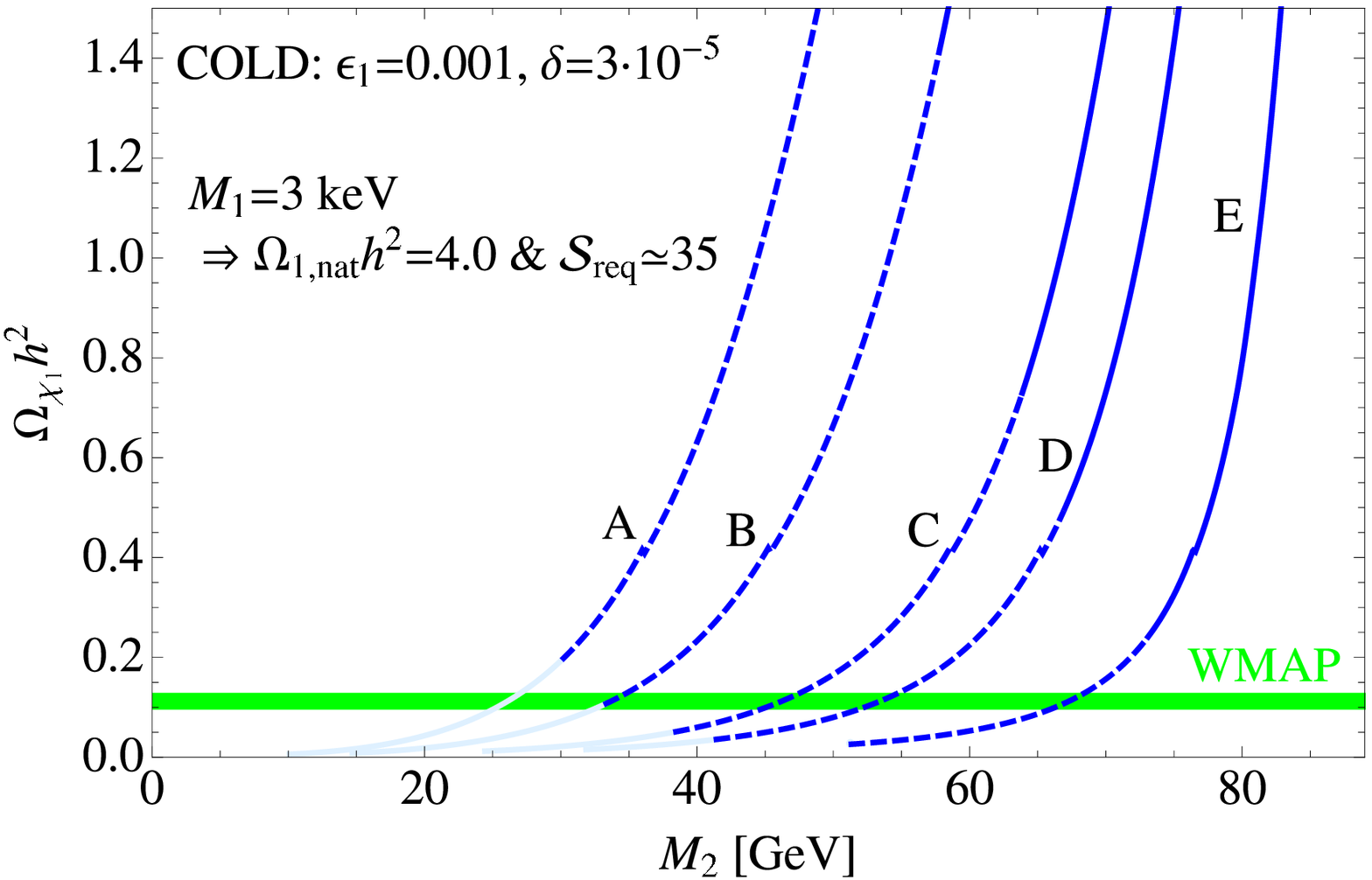}
\includegraphics[width=7.8cm]{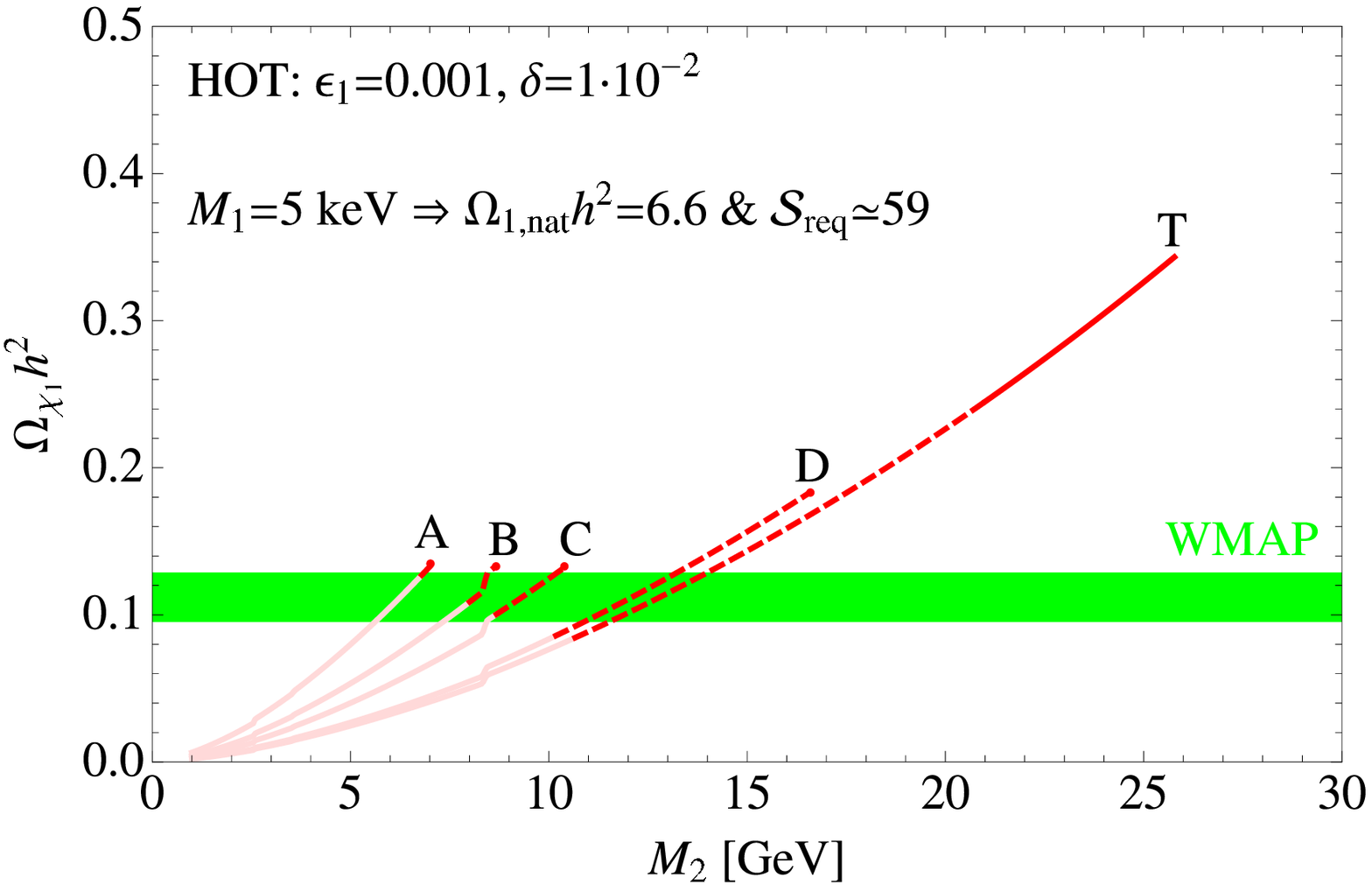}
\includegraphics[width=7.8cm]{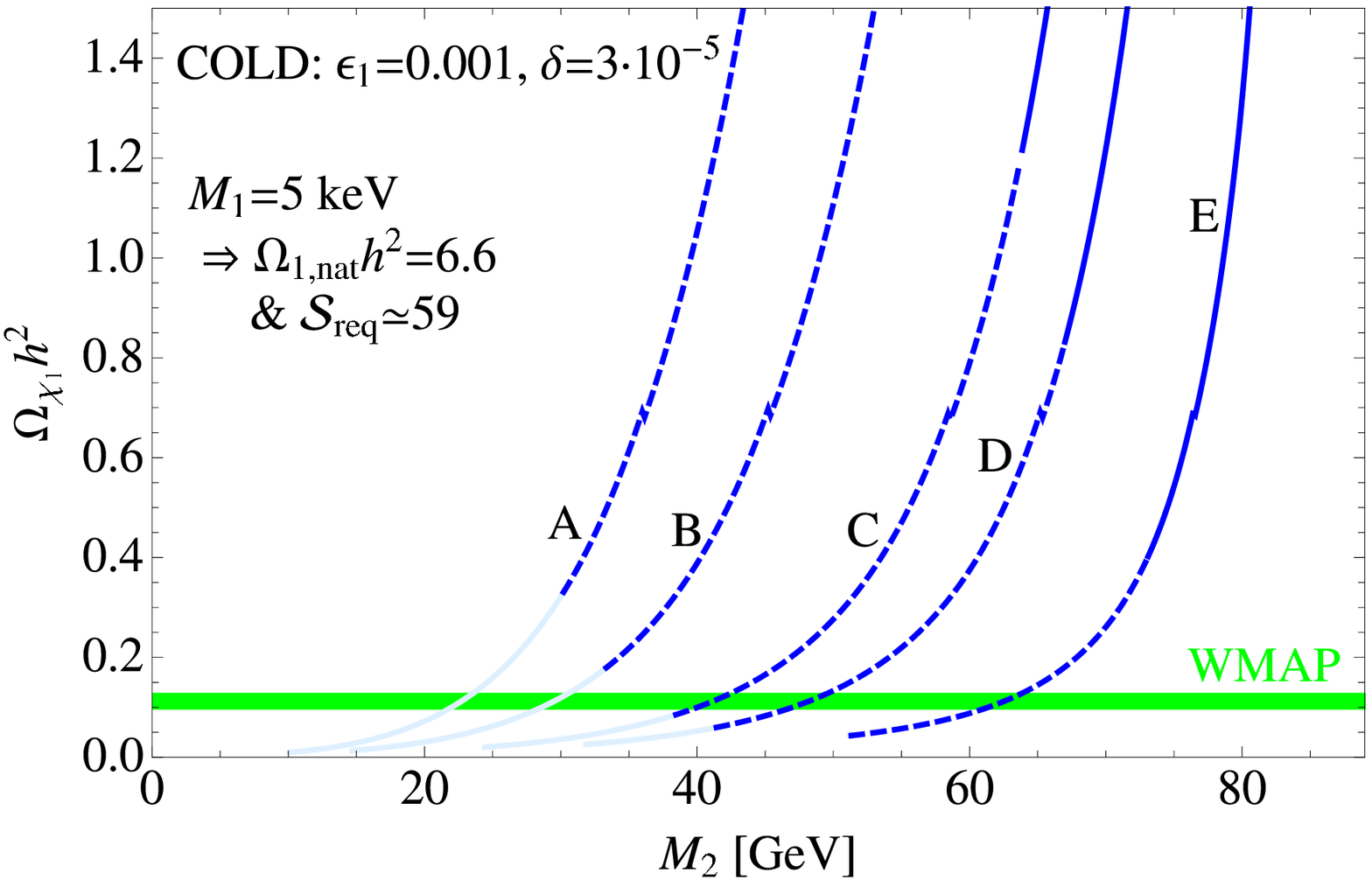}
\caption{\label{fig:Ab_Scen} Diluted DM abundances for the HOT and COLD scenarios. The colour code is the same as in Fig.~\ref{fig:Abun}. Note that the terms ``HOT'' and ``COLD'' refer to the freeze-out of $\chi_2$. The different curves correspond to those shown in Fig.~\ref{fig:Ent_Scen}.}
\end{figure}

Finally, the resulting diluted DM abundances for the two scenarios HOT and COLD are plotted in Fig.~\ref{fig:Ab_Scen}, for the two different example values of $M_1=3$~keV ($\simeq$ lower bound on $M_1$ for $\epsilon_1=0.001$) and $M_1=5$~keV ($>$ lower bound on $M_1$ for $\epsilon_1=0.001$). As indicated in the upper two panels, for a keVin with a mass close to the lower bound, one would need to correct the natural $\chi_1$ abundance by at least a factor of about 35, which is already hard to achieve within the range of the hard BBN bound. However, it is no problem if one only requires the weak BBN bound to be fulfilled. The lesson to learn is that the exact value of the bound is decisive for our framework, and it would be worthwhile to perform a detailed dedicated study. Requiring only the weak bound from now on, we can see from the plots that there is, especially for the COLD scenario, a sizable region in the parameter space, for which we can achieve the correct DM abundance. In particular, there is a certain trade-off between the parameters $M_1$, $M_2$, and $\epsilon_2$. Note that $M_1$ can in principle be freely chosen, as long as the lower bound from structure formation is not violated, cf.\ Fig.~\ref{fig:Lya}. However, increasing $M_1$ also increases the amount of entropy dilution, $\mathcal{S}_{\rm req}$, needed to obtain the correct DM abundance -- which will cut the amount of the $M_2$--$\epsilon_2$ space that leads to the correct abundance. In particular for the HOT scenario with $M_1=3$~keV (upper left panel of Fig.~\ref{fig:Ab_Scen}), we can see that there is a strong tendency to actually produce too much entropy for hot freeze-out of the $\chi_2$. In addition, this plot also shows that it would be worthwhile to study the intermediate freeze-out region of $\chi_2$, which cannot be done in a semi-analytical way. Still, this could very well be the ideally suited parameter region to obtain the correct abundance. We want to stress once more, however, that the true conclusion about this parameter region would require a more advanced numerical investigation of the Boltzmann equation.

One bonus of our general treatment is that we are able to investigate a large class of models within our framework. By comparing the interaction Lagrangians with the $Z$--boson with the general forms in Eqs.~\eqref{eq:Lii} and~\eqref{eq:L12}, we can express our general parameters in terms of model paramaters by the following mapping:
\begin{equation}
 (\epsilon_1, \epsilon_2, \delta, M_1, M_2) \to f({\rm model\ parameters}),
 \label{eq:mapping}
\end{equation}
where $f$ is a function that is characteristic for the concrete model under consideration. Hence, if we have a fully working set of parameters $(\epsilon_1, \epsilon_2, \delta, M_1, M_2)$, we can use Eq.~\eqref{eq:mapping} to obtain conditions on the model parameters. We have collected four such benchmark sets of parameters in Tab.~\ref{tab:benchmark}. All these points lead to DM abundances that are in agreement with the $3\sigma$ range for $\Omega_{\rm DM} h^2$ obtained by WMAP-7~\cite{Komatsu:2010fb}.

Note that, actually, it would be possible to generalize our considerations to models involving $Z'$--bosons, since this would in the zeroth order approximation only amount to a rescaling of the parameters $\epsilon_1$, $\epsilon_2$, and $\delta$. However, in addition to that, it may be that the $\chi_i$ coupling to the $Z$--boson is not of pure axial vector form, cf.\ Eqs.~\eqref{eq:Lii} and~\eqref{eq:L12}, and/or that the function $\Sigma$, cf.\ Eq.~\eqref{eq:MEs}, is modified, together with the replacement $(M_Z,\Gamma_Z) \to (M_{Z'},\Gamma_{Z'})$ in Eqs.~\eqref{eq:MEs}, \eqref{eq:sv_rel}, and~\eqref{eq:sv_non}. However, in situations where one is not interested in the region near the resonance peak, and where an $\mathcal{O}(1)$ approximation for $\Sigma$ is sufficient, one can still absorb most of the new model aspects in a rescaling of $(\epsilon_1, \epsilon_2, \delta)$.

In the following section, we will present some realistic cases where one could make use of our benchmark points.

\begin{table}[t]
\centering
\begin{tabular}{|c||c|c|c||c|c|c|}\hline
Benchmark & $\epsilon_1$ & $\delta$ & $M_1$ & $\chi_2$-FO & $\epsilon_2$ & $M_2$\\ \hline \hline
${\rm H}_{\rm D}$ & $0.001$ & $1\cdot 10^{-2}$ & $3$~keV & HOT   & $1.0\cdot 10^{-4}$ &\hfill $16$~GeV \\ \hline
${\rm H}_{\rm C}$ & $0.001$ & $1\cdot 10^{-2}$ & $5$~keV & HOT   & $1.5\cdot 10^{-4}$ &\hfill $9.5$~GeV \\ \hline \hline
${\rm C}_{\rm E}$ & $0.001$ & $3\cdot 10^{-5}$ & $3$~keV & COLD & $5.0\cdot 10^{-4}$ &\hfill $67$~GeV \\ \hline
${\rm C}_{\rm D}$ & $0.001$ & $3\cdot 10^{-5}$ & $5$~keV & COLD & $8.0\cdot 10^{-4}$ &\hfill $48$~GeV \\ \hline
\end{tabular}
\caption{\label{tab:benchmark} Four benchmark points that lead to the correct abundance and are all consistent with the Ly--$\alpha$ and the weak BBN bound. The classification ``HOT/COLD'' refers to the freeze-out of $\chi_2$, and the mass $M_2$ is chosen such that the correct DM abundance is obtained.}
\end{table}

\section{\label{sec:Concrete-keVins} Example Models}

After having discussed the general framework of keVins, we now want to look at concrete examples to illustrate in which types of models a keVin could arise as DM candidate. We discuss two realizations of our framework: first we will give an example of keVins within the $E_6$SSM, with the $Z$--boson being the dominant mediator of the $\chi_i$ annihilation. In that setup, the decay would involve a ``flavour'' changing coupling $\chi_2$ with $\chi_1$, and would hence be naturally suppressed compared to the annihilation channels, thereby justifying the use of our parameter $\delta$. Secondly, we will illustrate how to extend our considerations the case of keV sterile neutrinos. This extension needs additional constraints to be taken into account, which impose new difficulties not present in our general setting. However, when carefully adjusting our parameters, also this type of DM could be accommodated.

\subsection{\label{sec:E6SSM}A SUSY example: $E_6$SSM}

The situation of DM in the $E_6$SSM~\cite{King:2005jy,Hall:2011au,Athron:2011wu,Hall:2011zq} (see Ref.~\cite{Hall:2012yb} for a pedagogical introduction) is very similar to our general case presented in Sec.~\ref{sec:keVins}: we have one light mass eigenstate singlino $\chi^0_1$, which could play the role of a WDM particle, and a heavier brother of that particle, the next-to-lightest singlino $\chi^0_2$, which can decay into light particles. Both, $\chi^0_1$ and $\chi^0_2$, are thermally produced in the early Universe, and the decay of $\chi^0_2$ is responsible for the entropy production.

Our basic framework could actually be any supersymmetric version of the Standard Model, which will be denoted by SSM, as opposed to SM. Within this SSM framework, we need three ingredients:

\begin{enumerate}

\item Two (in general distinct) $SU(2)_L$ doublet superfields $\hat H_{1,2}$ with hypercharges $+1/2$ or $-1/2$. As the notation suggests, these might be the two Higgs doublet superfields of the minimal SSM (MSSM), or of any extended model.

\item Two distinct $SU(2)_L$ singlet superfields $\hat S_{1,2}$ with vanishing hypercharge. A generic example of such fields would be the singlets in the $E_6$SSM, which are usually referred to by a very similar notation as our singlets.

\item Some quantum number that distinguishes the fermionic components $s_{1,2}$ of $\hat S_{1,2}$ from right-handed (SM singlet) neutrinos. This job would usually be done by $R$--parity, which would in addition make the lightest inert mass eigenstate stable, although there could be other choices.\footnote{Note that this condition prevents the $s_{1,2}$ from mixing with neutrinos, which may by a big advantage or disadvantage of our setting, depending on the viewpoint: indeed, in the view of a model builder one avoids the strong constraint from the non-observation of the $\gamma$--line, while in the view of an observational astronomer one loses a striking signal.}

\end{enumerate}

In general, such a framework will tend to mix these gauge eigenstates (after electroweak symmetry breaking), which in particular applies to their fermionic components $h_i$ and $s_j$. We \emph{assume} this mixing to be such that, at tree-level, $h_i$ mixes with $s_i$ only by a small admixture of size $\epsilon_i$, while the mixing with $s_{j\neq i}$ could even be further suppressed, paramaterized by an additional factor of $\delta^{1/2}$. In such a situation, the mass eigenstates $\chi_{1,2,3,4}$ are given by\footnote{Note the order of the states on the left-hand side.}
\begin{equation}
 \begin{pmatrix}
 \chi_1\\
 \chi_3\\
 \chi_2\\
 \chi_4
 \end{pmatrix} \approx \begin{pmatrix}
  1 & \epsilon_1 & 0 & \xi^{1/2}\\
  -\epsilon_1 & 1 & \xi^{1/2} & 0\\
  0 & -\xi^{1/2} & 1 & \epsilon_2\\
  -\xi^{1/2} & 0 & -\epsilon_2 & 1
 \end{pmatrix} \begin{pmatrix}
 s_1\\
 h_1\\
 s_2\\
 h_2
 \end{pmatrix}.
 \label{eq:fermix}
\end{equation}
One should keep in mind that we are talking about a situation where the couplings to the $Z$--boson are suppressed, i.e., we have in mind that $\epsilon_i \ll 1$. Such a mixing pattern may look somehow engineered at first sight, but it might very well be motivated in a concrete model: in the $E_6$SSM, this structure suggests itself due to the third generation singlet $S_3$ being the only one to get a VEV. Note further that, since the matrix in Eq.~\eqref{eq:fermix} mixes $SU(2)_L$ singlets with doublets, it is natural for the corresponding couplings to be proportional to the electroweak VEV, $\epsilon_i \propto v$, which is normally small compared to any other VEVs. Furthermore, the ``flavour-changing'' vertices (i.e., $Z$--$\chi_1$--$\chi_2$ couplings) typically arise only at 1-loop level in the $E_6$SSM, which suggests an additional suppression parametrized by our parameter $\delta$ (or by $\xi= \epsilon_1 \epsilon_2 \delta$).

Due to their singlino nature, both $\chi^0_i$ have suppressed couplings to the $Z$. They are, however, charged under the full gauge group and couple with gauge strength to a heavier $Z'$ gauge boson. Depending on the mass of the $Z'$, these couplings could be suppressed. The best current lower limit on the mass of the $Z'$--boson has been obtained by the ATLAS collaboration~\cite{Collaboration:2011dca}, and it is given by $M_{Z'} > 1.83$~TeV. On the other hand, there is no a priori reason in the $E_6$SSM for the $Z'$ not to be even heavier, in which case the effective couplings of the $\chi^0_i$ to the $Z'$ are again very small. It is this limit that we will consider.

The decisive step is to compare the couplings in the model to the general forms given in Eqs.~\eqref{eq:Lii} and~\eqref{eq:L12}. Starting with the flavour diagonal couplings, one obtains~\cite{Hall:2011au}:
\begin{equation}
 g \epsilon_i^2 = \frac{M_Z}{2 v} R_{Zii},\ \ {\rm where}\ \ i=1,2.
 \label{eq:ExA_1}
\end{equation}
Here, $M_Z = \sqrt{g^2 + {g'}^2} v$ is the usual $Z$--boson mass, $R_{Zii} = \frac{v^2}{2 (m_{\chi_i^\pm})^2} \left( f_i^2 \cos^2 \beta - \tilde f_i^2 \sin^2 \beta \right)$ is the coupling parameter to the $Z$--boson, and $m_{\chi_i^\pm} = \frac{\lambda_i s}{\sqrt{2}}$ are the ``inert'' chargino masses. Furthermore, $s$ is the VEV of the third generation singlet scalar component $S_3$, $\tan \beta$ is the usual MSSM-like doublet VEV ratio, and the couplings $f_i$, $\tilde f_i$, and $\lambda_i$ arise from trilinear couplings of one singlino superfield to two Higgs doublet superfields in the $E_6$SSM superpotential~\cite{King:2005jy}.

Using Eq.~\eqref{eq:ExA_1}, one can express the general coupling $\epsilon_i$ in terms of model parameters:
\begin{equation}
 \epsilon_i = \frac{1}{\sqrt{2 c_W}} \sqrt{\omega_i^2 \cos^2 \beta - \tilde \omega_i^2 \sin^2 \beta}\ \frac{v}{s},
 \label{eq:ExA_2}
\end{equation}
where $\omega_i = f_i / \lambda_i$ and $\tilde \omega_i = \tilde f_i / \lambda_i$. Taking the ratio between $\epsilon_1$ and $\epsilon_2$ leads to
\begin{equation}
 \frac{\epsilon_1}{\epsilon_2} = \sqrt{\frac{\omega_1^2 - \tilde \omega_1^2 \tan^2 \beta}{\omega_2^2 - \tilde \omega_2^2 \tan^2 \beta}} \gg 1,
 \label{eq:ExA_3}
\end{equation}
as fulfilled for all of our four scenarios. Hence, although the VEV ratio $\frac{v}{s}$ determines the scale of the couplings $\epsilon_i$, one still needs some way to make the ratio in Eq.~\eqref{eq:ExA_3} large enough. We will discuss two possible solutions below.

Similarly, one can express the masses in terms of model parameters,
\begin{equation}
 M_i = \frac{\sin (2\beta)}{\sqrt{2}} \lambda_i \omega_i \tilde \omega_i\ \frac{v^2}{s}.
 \label{eq:ExA_4}
\end{equation}
Like before, the size of the masses is determined by $\frac{v^2}{s}$, which suggests that both are significantly smaller than the electroweak VEV $v$. Taking again the mass ratio,
\begin{equation}
 \frac{M_1}{M_2} = \frac{\lambda_1 \omega_1 \tilde \omega_1}{\lambda_2 \omega_2 \tilde \omega_2} \ll 1.
 \label{eq:ExA_5}
\end{equation}
In order to simultaneously fulfill Eqs.~\eqref{eq:ExA_3} and~\eqref{eq:ExA_5}, we have different possibilities:
\begin{itemize}

\item \underline{flipped hierarchies}: $f_1 > f_2, \tilde f_2 \gg \tilde f_1$\\
If $\lambda_1 \approx \lambda_2$, then this condition leads immediately to $M_1 \ll M_2$, due to $\tilde f_1 \ll \tilde f_2$. In turn, Eq.~\eqref{eq:ExA_3} simplifies to
\begin{equation}
 \frac{\epsilon_1}{\epsilon_2} \approx \frac{f_1}{\sqrt{f_2^2 - \tilde f_2^2 \tan^2 \beta}},
 \label{eq:ExA_6}
\end{equation}
whose numerator is much larger than the denominator.

\item \underline{tuned $\tan \beta$}: $\tan \beta \simeq f_2/\tilde f_2$\\
This assumption leads to $\omega_2^2 - \tilde \omega_2^2 \tan^2 \beta \approx 0$ in Eq.~\eqref{eq:ExA_2}, and hence to a very small coupling $\epsilon_2$. However, this condition still allows all other combinations of couplings to be sizable. Thus, in order to also fulfill the condition in Eq.~\eqref{eq:ExA_5}, it is necessary to impose an additional assumption that leads to $f_1 \tilde f_1 / \lambda_1 \ll f_2 \tilde f_2 / \lambda_2$.

\end{itemize}

We still have to test our original assumption of a $Z'$--boson that us practically decoupled: $M_2/v \sim 0.01\textrm{--}0.1$, together with Eq.~\eqref{eq:ExA_4}, suggests that $s \sim \mathcal{O}(10 \textrm{--} 100)\ v$. Hence, a typical annihilation cross section involving a $Z'$--boson instead of a $Z$--boson would be suppressed by a factor of about $\left(M_Z / M_{Z'} \right)^4$, which is approximately $\left( v / s \right)^4 \lesssim 10^{-4}$. This has to be compared to the natural $\epsilon_i^4$ suppressions of the cross sections involving the ordinary $Z$--boson, in order to ensure that the $Z'$--contribution is indeed subdominant. For $\epsilon_2 \sim 10^{-4}$, this leads to the condition
\begin{equation}
 M_{Z'} \sim s \gg \frac{v}{\epsilon_2} \gtrsim \mathcal{O} \left( 10^{3 \textrm{--} 4} \right) v.
 \label{eq:ExA_7}
\end{equation}
But this and Eq.~\eqref{eq:ExA_4} would lead to $M_2 \lesssim \frac{v}{s}\ v \lesssim 10^{-3} v \sim 0.1$~GeV, which is too small to be consistent with the reheating bound, cf.\ Sec.~\ref{sec:keVins}. A simple way out would be to assume that the $Z'$--boson mass receives additional contributions from further new physics, thereby invalidating the above assumption $M_{Z'} \sim s$. Then we could have a much stronger suppression of the $Z'$--mediated couplings, justifying the small values for the parameter $\delta$ from Tab.~\ref{tab:benchmark}, cf.\ Eq.~\eqref{eq:L12}. This off-diagonal coupling arises in the $E_6$SSM only at loop level, as it vanishes at tree-level for $\lambda_\alpha s \gg f_{\alpha} v,\ \tilde f_{\alpha} v$~\cite{Hall:2009aj}. Even though we cannot give an analytical formula, one could generically expect a size of
\begin{equation}
 R_{Z12} \sim \sqrt{R_{Z11} R_{Z22}} \tilde \delta,
 \label{eq:ExA_8}
\end{equation}
where $\tilde \delta$ is some suppression factor. Comparing this to Eq.~\eqref{eq:L12} leads to
\begin{equation}
 \epsilon_1 \epsilon_2 \delta = \frac{M_Z}{2 v} R_{Z12} \tilde \delta,
 \label{eq:ExA_9}
\end{equation}
which translates into $\delta =\frac{1}{2 c_W} \ \tilde \delta$. Hence, $\tilde \delta$ should be tuned to have a somehow similar value as $\delta$ in the benchmark scenario under consideration. The natural suppression factor for a loop would be about $\tilde \delta \sim 1/(16 \pi^2)$, which is okay with scenarios ${\rm H}_{\rm D,C}$ from Tab.~\ref{tab:benchmark}, while ${\rm C}_{\rm E,D}$ would require some more suppression. That could come from the structure of the loop which could exhibit, e.g., a classical~\cite{Glashow:1970gm} or extended~\cite{Blum:2007he} GIM mechanism.

Trying to find explicit realizations for the different benchmark points from Tab.~\ref{tab:benchmark}, it is easiest to perform a $\chi^2$--minimization. Good fits were found for scenarios ${\rm H}_{\rm D,C}$ and ${\rm C}_{\rm D}$, and they are listed in Tab.~\ref{tab:ExA}. Note that the fits for ${\rm H}_{\rm C}$ and ${\rm C}_{\rm D}$ both correspond to the tuned $\tan \beta$ case mentioned above, while the one for ${\rm H}_{\rm D}$ corresponds to the case of flipped hierarchies.

\begin{table}[t]
\centering
\begin{tabular}{|c||c|c|c|c|}\hline
  & ${\rm H}_{\rm D}$ & ${\rm H}_{\rm C}$ & ${\rm C}_{\rm E}$ & ${\rm C}_{\rm D}$\\ \hline \hline
  $f_1$ & $2.4\cdot 10^{-2}$ & $6.1\cdot 10^{-3}$ & ------------ & $2.0\cdot 10^{-3}$ \\
  $\tilde f_1$ & $1.4\cdot 10^{-5}$ & $1.3\cdot 10^{-5}$ & ------------ & $1.3\cdot 10^{-5}$ \\ \hline
  $f_2$ & $3.5\cdot 10^{-1}$ & $1.0$ & ------------ & $1.0$ \\
  $\tilde f_2$ & $6.8\cdot 10^{-1}$ & $0.67$ & ------------ & $0.67$ \\ \hline
  $\lambda_1$ & $1.1$ & $1.0\cdot 10^{-1}$ & ------------ & $8.4\cdot 10^{-2}$ \\
  $\lambda_2$ & $1.4\cdot 10^{-1}$ & $0.43$ & ------------ & $0.23$ \\ \hline
  $\tilde \delta$ & $1.8\cdot 10^{-2}$ & $1.8\cdot 10^{-2}$ & ------------ & $5.3\cdot 10^{-5}$ \\
  $\tan \beta$ & 0.52 & $1.5$ & ------------ & $1.5$ \\
  $s$ [TeV] & $3.78$ & $6.37$ & ------------ & $2.38$ \\ \hline 
\end{tabular}
\caption{\label{tab:ExA} Parameter values for the $E_6$SSM with approximate decoupling of the $Z'$--boson. Note that we have not been able to find a fitting parameter combination for ${\rm C}_{\rm E}$.}
\end{table}

\subsection{\label{sec:SterileNus}keV sterile neutrinos in a Left-Right symmetric context}

In this section, we would like to shortly illustrate that it is, in principle, possible to extend our considerations to the case of keV sterile neutrino Dark Matter, where the lightest sterile neutrino $N_1$ is assumed to have a keV-scale mass. However, this extension also introduces additional constraints have to be taken into account, in particular the astrophysical bound on the radiative decay $N_1\to \nu \gamma$~\cite{Watson:2006qb,Yuksel:2007xh} and constraints from active-sterile neutrino mixing. The idea of using additional entropy production to dilute the thermal abundance of keV sterile neutrinos was, to our knowledge, first mentioned in Ref.~\cite{Asaka:2006ek}. The key point is to assume that the $N_1$, although being a SM singlet, shares some interaction at high energies. A very concrete example of such a case has been worked out in Ref.~\cite{Bezrukov:2009th}, where a Left-Right ($LR$) gauge symmetry was assumed under which the right-handed (SM-singlet) neutrinos are charged. We will refer here to the notation used in Ref.~\cite{Bezrukov:2009th} to give a flavour of what would change when extending our considerations.

To generalize our considerations to the case of sterile neutrinos in an $LR$-symmetric framework, we first make the identification $(\chi_1, \chi_2) \to (N_1, N_2)$. Then, one immediately observes that the corresponding sterile neutrino masses $M_1$ and $M_2$ are free parameters, and can be tuned to have, e.g., the values of one of our benchmark scenarios in Tab.~\ref{tab:benchmark}.\footnote{Note that this statement is only true as long as there is no reason for the masses to have certain values in a particular model. There is still not too much work done in that direction, but among the possibilities to explain such a mass pattern involving a strong splitting are the \emph{split seesaw} mechanism~\cite{Kusenko:2010ik} (which can be supplemented by an $A_4$ symmetry~\cite{Adulpravitchai:2011rq}), by $L_e - L_\mu -L_\tau$ symmetry~\cite{Shaposhnikov:2006nn,Lindner:2010wr} (or other flavour symmetries~\cite{Araki:2011zg}), by the \emph{Froggatt-Nielsen} mechanism~\cite{Barry:2011wb,Merle:2011yv,Barry:2011fp}, or by the so-called \emph{extended seesaw}~\cite{Zhang:2011vh}, as mentioned in Sec.~\ref{sec:intro}.} Reproducing the correct values of the paramaters $(\epsilon_1, \epsilon_2, \delta)$ is a bit less trivial. Still, it is possible to redefine our paramaters in terms of model paramaters: although the corresponding decay and annihilation processes are mediated partially by heavy charged $W_R^\pm$ bosons, the structure of the diagrams does not change considerably. Because of the pure gauge interaction, we need $\epsilon_1 = \epsilon_2$, due to the gauge interaction, which is not possible in any of our scenarios. However, it is easy to find other points in parameter space where the correct abundance is reproduced. One such point with $\epsilon_1 = \epsilon_2$ would be the following:
\begin{equation}
 (\epsilon_1, \epsilon_2, \delta, M_1, M_2) = (10^{-4}, 10^{-4}, 1, 2.6~{\rm keV}, 4.4~{\rm GeV}),
 \label{eq:sterile_nu_scen}
\end{equation}
which produces an abundance of $\Omega_{N_1} h^2 = 0.113$ and a reheating temperature that is in accordance with the weak BBN bound. In addition, the keVin mass of $M_1 = 2.6~{\rm keV}$ is close to but still above the Ly--$\alpha$ bound.

Taking this new benchmark point, we can compare the suppressed coupling to weak interaction from Eq.~\eqref{eq:Lii} to the strength of the suppressed coupling of $N_1$ to the heavy right-handed $W_R$--bosons used in Ref.~\cite{Bezrukov:2009th}. By comparing the cross sections, and disregarding the differences between the $\gamma_5$ and the $P_R$ coupling for simplicity, we obtain
\begin{equation}
 \epsilon_{1,2} = 0.0001 \approx 1.27\ \sqrt{G_F c_W} \frac{M_W M_Z}{M_{W_R}},
 \label{eq:ExC_1}
\end{equation}
leading to $M_{W_R} = 2.97\cdot 10^5$~GeV. This scale looks much smaller than the value of $v_R \sim 10^5$~TeV quoted in Ref.~\cite{Bezrukov:2009th}, so apparently the numerical example given there does not correspond our benchmark point. This is, however, not too much of a wonder since we did not attempt to find all working parameter combinations, but rather wanted to give a proof-of-principle. We can go one step further by comparing our approximate decay width, Eq.~\eqref{eq:G2_approx}, with the estimate $\Gamma_{N_2} \gtrsim \frac{G_F^2 M_2^5}{192 \pi^2} \ \theta_2^2$~\cite{Bezrukov:2009th} and again disregarding the $\gamma$--structure, which leads to a second generation active-sterile mixing angle of $\theta_2 \approx 6.9\cdot 10^{-8}$, in good accordance with the value obtained in Ref.~\cite{Bezrukov:2009th}. Note, however, that their scenario is somehow less restricted than ours, since the authors assume the gauge interactions to dominate the DM production, while they take the Yukawa couplings to be mainly responsible for the decay of $N_2$. This disentangles certain requirements, which in our case can only be partially resembled by choosing a small value for $\delta$.

However, as already mentioned, additional constraints come into the game, the strongest being in this case the upper bound on the radiative decay width, $\Gamma (N_1\to \nu \gamma)$, which arises from the non-observation of the corresponding monoenergetic X-ray line~\cite{Watson:2006qb,Yuksel:2007xh}.\footnote{Note that additional signals, like e.g.\ enhanced dipole moments~\cite{Geng:2012jm}, might show up in a cosmological context.} According to Ref.~\cite{Bezrukov:2009th}, this bound would constrain the mixing angle $\theta_1$ to be less than about $3.8\cdot 10^{-4}$ for $M_1 = 2.6~{\rm keV}$. On the other hand, active-sterile neutrino mixing constrains a certain combination of masses and mixing angles to be larger than the square root of the solar neutrino mass square difference,
\begin{equation}
 M_1 \theta_1^2 + M_2 \theta_2^2 > \sqrt{8\cdot 10^5}~{\rm eV} \simeq 0.009~{\rm eV},
 \label{eq:ExC_2}
\end{equation}
supposed that we have a type~I seesaw situation. It is exactly this bound, Eq.~\eqref{eq:ExC_2}, which cannot be fulfilled with the masses and mixing angles obtained above, in agreement with the findings in Ref.~\cite{Bezrukov:2009th}. Hence, indeed we have seen that it is the \emph{additional} constraints that generically arise in the framework of keV sterile neutrinos, which could destroy the validity of an otherwise working benchmark point. Thus one has to be careful when extending our findings to this case, since additional strong bounds might play a decisive role. However, the constraint from Eq.~\eqref{eq:ExC_2} can be avoided in a seesaw type~II situation, which naturally arises in $LR$-symmetric extensions of the SM. This possibility has also been worked out in detail in Ref.~\cite{Bezrukov:2009th} and has the feature of being much more model-dependent, due to all the detailed considerations to be undergone when attempting to properly take into account neutrino mixing.

Let us end by noting that, as mentioned in Sec.~\ref{sec:keVins_numerical}, the amount of entropy produced could be increased by involving more than one generation of decaying fermions. In the case of a keV sterile neutrino, this possibility is particularly attractive, as there would normally be yet another heavier sterile neutrino $N_3$ which could also contribute to the entropy production. In particular in the context of $LR$-symmetry this particle must exist, and one could use it to dilute the $N_1$ abundance even further, if required.

So, indeed, it is possible to extend our general consideration to the case of keV sterile neutrinos, but the situation is more complicated and leads to a strong loss of generality, the latter being the main reason for us not to focus on the special case of keV sterile neutrinos here.

\section{\label{sec:conclusions}Conclusions}

In this paper, we have proposed a simple model for Warm Dark Matter (WDM) in which two fermions are added to the Standard Model: stable ``keVins'' ( keV inert fermions) $\chi_1$ which account for WDM and their unstable brothers,  the ``GeVins'' ( GeV inert fermions) $\chi_2$, both of which carry zero electric charge and lepton number, and are (approximately) ``inert'' in the sense that their only interactions are via suppressed couplings to the $Z$. We have considered scenarios in which stable keVins are thermally produced and their abundance is subsequently diluted by entropy production from the decays of the heavier unstable GeVins. 

We have investigated in detail the question under which conditions it is possible to arrive at the correct relic abundance with these particles, which is done in our case by thermal overproduction and subsequent dilution of the abundance by additional entropy production in the decay of the $\chi_2$ GeVin, which is assumed to have a mass around the GeV scale in our benchmark scenarios. We have identified four example benchmark scenarios for which it is possible to achieve the correct abundance, depending on the nature of the freeze-out of $\chi_2$. 

This mechanism could be implemented in a wide variety of models, including $E_6$ inspired supersymmetric models in which a keVin candidate naturally appears, or models involving sterile neutrinos. For example, in the $E_6$ models the keVins are mainly Standard Model singlet fermions with only very small active admixtures due to mass basis rotations allowing them to have strongly suppressed couplings to the $Z$--boson.

Although we have been able to give a proof-of-principle in this work, we nevertheless have applied simplifications on several places, in particular in what concerns the freeze-out processes. The natural subsequent step would be to perform a more elaborate numerical study of the Boltzmann equation in order to extend our considerations to the intermediate freeze-out region of $\chi_2$, and by this to directly calculate the freeze-out temperatures. Further investigations should be done, e.g., by identifying further regions in parameter space that lead to the correct relic abundance, and by using them to map other concrete models onto our general framework. We hope that we have been able to perform the first step in this direction and that many future works will investigate the framework of keVins in great detail.

\section*{Acknowledgments}

We would like to thank F.~Bezrukov and P.~di Bari for useful discussions. 
SFK acknowledges partial support 
from the STFC Consolidated ST/J000396/1 and EU ITN grants UNILHC 237920 and INVISIBLES 289442 .
The work of AM is supported by the G\"oran Gustafsson foundation.

\bibliographystyle{./apsrev}
\bibliography{./keVins}

\end{document}